\documentclass[%
 reprint,
showpacs,preprintnumbers,
 amsmath,amssymb,
 aps,
]{revtex4-1}

\usepackage{subfigure}
\usepackage[rflt]{floatflt}
\usepackage{color}

\usepackage{graphicx}
\usepackage{dcolumn}
\usepackage{bm}

\begin{document}


\title{Thermodynamic Properties of the Anisotropic Frustrated Spin-chain Compound Linarite PbCuSO$_4$(OH)$_2$}%

\author{M.\ Sch\"{a}pers$^1$, A.U.B.\ Wolter$^1$, S.-L.\ Drechsler$^1$, S.\ Nishimoto$^1$, K.-H.\ M\"uller$^1$, M.\ Abdel-Hafiez$^1$, W. Schottenhamel$^1$, B.\ B\"{u}chner$^{1,5}$,  J.\ Richter$^2$, B. Ouladdiaf$^3$, M.\ Uhlarz$^4$, R.\ Beyer$^{4,5}$, Y.\ Skourski$^4$, J.\ Wosnitza$^{4,5}$, K.C.\ Rule$^{6,7}$, H.\ Ryll$^6$, B.\ Klemke$^6$, K.\ Kiefer$^6$, M. Reehuis$^6$, B.\ Willenberg$^{6,8}$, S.\ S\"{u}llow$^8$}

\address{$^1$Leibniz Institute for Solid State and Materials Research IFW Dresden, D-01171 Dresden, Germany\\
$^2$Institute for Theoretical Physics, University of Magdeburg, D-39016 Magdeburg, Germany\\
$^3$Institute Laue-Langevin, F-38042 Grenoble Cedex, France\\
$^4$Dresden High Magnetic Field Laboratory, Helmholtz-Zentrum Dresden-Rossendorf, D-01314 Dresden, Germany\\
$^5$Institut f\"{u}r Festk\"{o}rperphysik, TU Dresden, D-01069 Dresden, Germany\\
$^6$Helmholtz Center Berlin for Materials and Energy, D-14109 Berlin, Germany\\
$^7$The Bragg Institute, ANSTO, Kirrawee DC NSW 2234, Australia\\
$^8$Institute for Physics of Condensed Matter, TU Braunschweig,
D-38106 Braunschweig, Germany}

\date{\today}

\begin{abstract}
We present a comprehensive macroscopic thermodynamic study of the
quasi-one-dimensional (1D) $s = \tfrac{1}{2}$ frustrated
spin-chain system linarite. Susceptibility, magnetization,
specific heat, magnetocaloric effect, magnetostriction, and
thermal-expansion measurements were performed to characterize the
magnetic phase diagram. In particular, for magnetic fields along
the $b$ axis five different magnetic regions have been detected,
some of them exhibiting short-range-order effects. The
experimental magnetic entropy and magnetization are compared to a
theoretical modelling of these quantities using DMRG and TMRG
approaches. Within the framework of a purely 1D isotropic model
Hamiltonian, only a qualitative agreement between theory and the
experimental data can be achieved. Instead, it is demonstrated
that a significant symmetric anisotropic exchange of about 10\,\%
is necessary to account for the basic experimental observations,
including the 3D saturation field, and which in turn might
stabilize a triatic (three-magnon) multipolar phase.
\end{abstract}

\maketitle


\section{Introduction}
Within the last four decades modern research on magnetic materials
has focussed on studying low-dimensional (quantum) spin systems
\cite{Lake2005,Sebastian2006}. From such investigations, these
compounds have been found to possess exotic physical ground-state
properties such as resonating valence bond \cite{Anderson1987},
quantum spin liquid \cite{Han2012}, and spin Peierls ground states
\cite{Hase1993}.

Nearly one-dimensional (1D) coupled quantum magnets can be
realized, for instance, in chain-like arrangements of spins of $s
= \tfrac{1}{2}$ Cu$^{2+}$ or V$^{4+}$ cations, that are typically
surrounded by oxygen anions. In general, the basic building blocks
of a Cu-oxide spin-chain system are CuO$_4$ plaquettes which are
connected to each other along one crystallographic direction,
viz., one dimension. Here, we focus on this type of copper oxides,
where one needs to distinguish between two different classes of
materials. In one class of compounds the linkage along the chain
occurs at the corners of the plaquettes, thus forming the
so-called corner-sharing chain. This geometrical configuration
leads to a linear Cu-O-Cu bond between neighboring Cu ions. Then,
the oxygen 2$p$ orbitals hybridize with the copper 3$d$ orbitals
with a straight bond angle of 180$^{\circ}$, hence the
Goodenough-Kanamori-Anderson rules predict a strong
antiferromagnetic (AFM) exchange interaction along the chain
between all nearest-neighbor (NN) Cu ions resulting essentially in
an unfrustrated system. These systems can be described to the
first approximation by the now reasonably well understood simple
AFM Heisenberg models extensively studied theoretically for more
than eighty years.

In contrast, a second class of compounds contains edge-sharing
CuO$_4$ units. In this situation the bond angle between the
nearest Cu-ion neighbors (NN), Cu-O-Cu, is close to 90$^{\circ}$,
which leads in most cases to a ferromagnetic coupling (FM) along
this bond. The AFM superexchange contribution is very weak for
such a geometry according to the Goodenough-Kanamori-Anderson
rules, since it vanishes exactly in the case of a 90$^{\circ}$
Cu-O-Cu bond angle. Under such circumstances the dominant FM $J_1$
stems mainly from the relatively direct large FM interaction
$K_{pd} \approx 900$\,K between holes on neighboring oxygen and
copper sites \cite{Mizuno1999,Lorenz2009,Kuzian2012} and not from
the Hunds coupling between the mentioned two oxygen orbitals as
frequently believed. The latter contributes about 20\,\% to the
value of $J_1$, only. In comparison, the next-nearest-neighbor
(NNN) Cu-O-O-Cu exchange paths contain $\sigma$ bonds of oxygen
$2p$ orbitals resulting in an AFM coupling which always causes
frustration effects, irrespective of the sign of the NN coupling
and in particular it is almost independent of $K_{pd}$, in sharp
contrast to $J_1$ which exhibits a very sensitive linear
dependence on $K_{pd}$ \cite{Malek2013}. Comparable to the first
case, in this second class of compounds the NN and NNN
interactions are often similar in magnitude leading to strong
frustration which offers a large variety of possible ground
states. The scientific history of this class and the related
quantum models are much younger (tracing back to the last decade)
than that of the simpler well-investigated AFM Heisenberg $s =
\tfrac{1}{2}$ chain.

\begin{figure}
\begin{center}
\includegraphics[width=0.9\columnwidth]{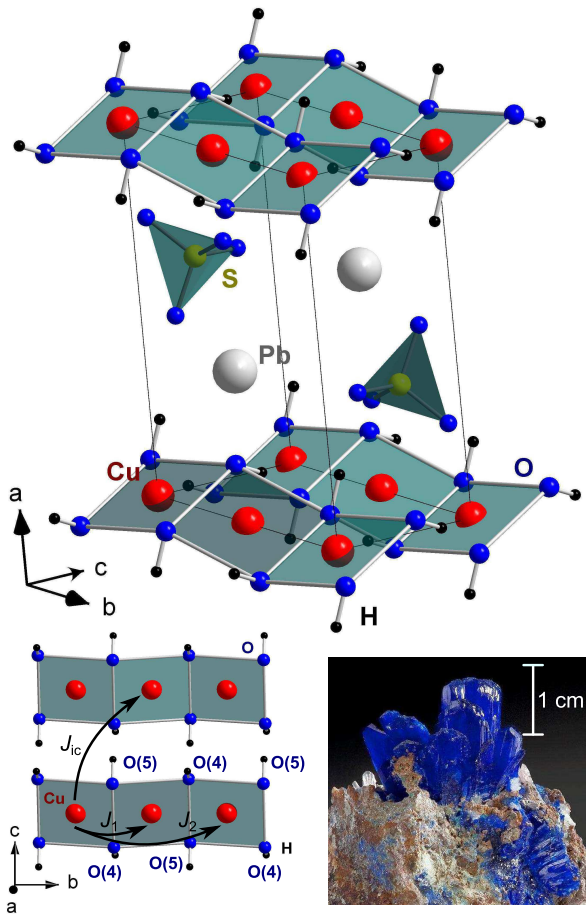}
\end{center}
\caption{(Color online) Upper part: The crystallographic structure
of PbCuSO$_4$(OH)$_2$ consisting of buckled neutral Cu(OH)$_2$
chains propagating along the crystallographic $b$ direction
surrounded by Pb$^{2+}$ cations and SO$_4^{2-}$ anions. Lower
panel, left: The main exchange paths $J_1$ and $J_2$ (notation in the general anisotropic case for the two intrachain exchange paths shown would be $\Delta_1J_1$ and $\Delta_2J_2$, see Eq.~\ref{Hamiltonian} and the text below) in the basal $bc$ plane as well as
the dominant skew interchain coupling $J_\text{ic}$. The photographic
picture shows one of our mineral specimens from the Grand Reef
Mine in Graham County, Arizona.} \label{linarite}
\end{figure}

Various Cu-oxide materials have been discovered which represent
excellent experimental realizations of such quasi-1D quantum
magnets (Q1DQM), e.g., LiCuVO$_4$ \cite{Enderle2005},
LiCu$_2$O$_2$ \cite{Gippius2004,Park2007}, Li$_2$ZrCuO$_4$
\cite{Drechsler2007}, and LiCuSbO$_4$ \cite{Dutton2012}. The basic
model to describe the interplay of the NN and NNN exchange for the
magnetic properties is the so-called 1D isotropic $J_1$-$J_2$ or
zig-zag chain (ladder) model, which corresponds to the Hamiltonian

\begin{equation}
\begin{split}
\hat{H} &= J_1 \sum_{l} {\bf S}_l\cdot{\bf S}_{l+1} + J_2
\sum_{l} {\bf S}_l\cdot{\bf S}_{l+2} +\\
&+ \sum_{l} \left( D_1-1 \right)J_1{ S}^z_l{
S}^z_{l+1} + \left( D_2-1 \right)J_2{ S}^z_l{ S}^z_{l+2} +\\
&-h \sum_{l} S_l^z
\end{split}
\label{Hamiltonian}
\end{equation}

Here, $J_1 < 0$ is the FM NN-interaction, $J_2 > 0$ is the AFM NNN
exchange, and $h=g \mu_B H$ represents the external magnetic field
along the easy ($z$) direction. The symmetric exchange anisotropy
\cite{remarkM} terms with the anisotropy parameters $D_{1,2}$ to
be discussed  in Sec.~V are given in the second line of
Eq.~\ref{Hamiltonian}. Depending on the frustration ratio $\alpha
= -J_2 / J_1$ and within the limits of a classical approach with
isotropic exchange, theory predicts various ground states for this
class of materials: For an $\alpha$ value $0 < \alpha <
\tfrac{1}{4}$ a FM ground state should occur, a value between
$-\tfrac{1}{4} < \alpha < 0$ should result in a collinear AFM
N\'{e}el ground state, while for all other values a non-collinear
spin-spiral ground state is predicted
\cite{Hamada1988,Tonegawa1989}.

If we also consider weak interchain interaction, anisotropic
couplings and quantum fluctuations, which may actually strongly
affect the 3D magnetic ordering, theory predicts even more exotic
ground states \cite{Furukawa2010}. Moreover, by applying an
external magnetic field a rich variety of exotic field-induced
phases may occur in these materials
\cite{Hikihara2008,Sudan2009,Zhitomirsky2010}. The recent
discovery of multiferroicity in LiCu$_2$O$_2$
\cite{Park2007,Seki2008} and LiCuVO$_4$
\cite{Naito2007,Schrettle2008,Mourigal2011}, as predicted by
theory \cite{Katsura2005,Sergienko2006,Mostovoy2006,Xiang2007} for
spin-chain systems with a helical ground state, has opened up
another playground in this research area. Unfortunately, the
Li$^+$ ions tend to interchange with the Cu$^{2+}$ ions in the
aforementioned materials, therefore, the microscopic source for
multiferroicity has not yet been established
\cite{Moskvin2008,Moskvin2009}.

Consequently, in order to experimentally investigate these
different phenomena and ground states, a material is required
which ideally exists in single crystal form without positional
disorder, exhibits anisotropic exchange, and possesses a
saturation field that is within experimental reach. In a recent
investigation we have shown \cite{Wolter2012} that the natural
mineral linarite, PbCuSO$_4$(OH)$_2$, satisfies all of these
requirements, thus offering unique possibilities to study a
variety of the above-mentioned physical topics.

Linarite crystallizes in a monoclinic lattice (space-group
symmetry $P2_1/m$; $a = 9.682$\,\AA, $b = 5.646$\,\AA, $c =
4.683$\,\AA, $\beta$ = $102.65^{\circ}$ \cite{Schofield2009}). In
linarite the chains are formed by Cu(OH)$_4$ units connected along
the $b$ direction in a buckled, edge-sharing geometry. In a
previous study \cite{Wolter2012} the $b$ direction was found to be
the easy axis of the system. Consequently, the Cu$^{2+}$ ions
($3d^9$ configuration) form an $s = \tfrac{1}{2}$ quasi-1D spin
chain along the $b$ direction (illustrated in
Fig.~\ref{linarite}), since the distance between two neighboring
Cu ions along the $b$ direction is much smaller than along the
other crystallographic directions. The surrounding oxygen orbitals
mediate the main exchange between the spins residing on the Cu
ions along the chain. As explained above, the $J_1$ is FM and the
largest coupling in the whole system. Due to the competition
between that FM NN and the AFM NNN exchange linarite has been
established as a magnetically frustrated system. Each oxygen atom
binds a hydrogen atom, whereas in between the chains one SO$_4$
tetrahedron and one lead atom complete the elemental unit cell.
The latter act as spacers between the chains and are responsible
for its quasi-1D nature.

A recent detailed study of the paramagnetic regime of linarite
revealed the coupling constants to be $J_1 \approx -100$\,K and
$J_2 \approx 36$\,K \cite{Wolter2012}. In effect, a frustration
ratio $\alpha = -J_2/J_1 \approx 0.36$ is found, which is much
closer to the 1D critical point ($\alpha = 0.25$) as compared to
the values reported in earlier studies \cite{Baran2006,Yasui2011}.
Because of a finite interchain coupling the system undergoes a
transition into a long-range magnetically ordered state below $T
\approx 2.8$\,K. The magnetic ground state was found to consist of
an elliptical helical structure with an incommensurate propagation
vector ${\bf k} = (0,\,0.186,\,0.5)$ \cite{Willenberg2012}.

Here, we present an extensive study of the physical properties of
linarite in zero and applied magnetic fields. We will show that
relatively weak magnetic fields of a few Tesla have a significant
influence on the physical properties of linarite and on the
low-temperature specific heat, in particular. This behavior can
qualitatively be explained within the framework of the model of a
NN-NNN frustrated spin chain, if other terms such as exchange
anisotropy are included in the Hamiltonian. The paper is organized
as follows: First, we present an extensive study of the
low-temperature thermodynamics in zero and applied field of
single-crystalline linarite. From the data we establish the
magnetic phase diagram for the three crystallographic directions.
Finally, we discuss our data, in particular in context of
numerical modelling approaches based on one-dimensional spin
models and extensions to these.

\section{Experimental}

\subsection{Samples and diffraction}

The single crystals of PbCuSO$_4$(OH)$_2$ used in this study are
natural minerals from different sources. In Tab.
\ref{tab:crystals}, a summary is presented on the use of the
different crystals for the set of experimental methods employed in
this work. All crystals show well-defined facets and the principal
axes $b$ and $c$ can be identified easily. With these also the
normal to the $bc$ plane, $a_{\perp}$, is determined (see
Ref.~\cite{Wolter2012} for this particular choice of crystal
direction). The crystal quality of our samples have been checked
by Laue X-ray diffraction. For all sets of single crystals no
magnetic impurity phases were observed within experimental
resolution, as evidenced by the absence of a low-temperature Curie
tail in the magnetic susceptibility. For all measurements the
samples were oriented along the crystallographic directions
$a_{\perp}$, $b$, and $c$ with a possible misalignment of less
than $5^{\circ}$.

\subsection{Susceptibility and magnetization}

In the $^4$He temperature range, the DC susceptibility was
measured by using a commercial vibrating sample magnetometer
(VSM). Magnetization measurements for magnetic fields along
$a_{\perp}$, $b$, and $c$ at fixed temperatures between 1.8 and
2.8\,K have been performed using a Physical Properties Measurement
System (PPMS) with a VSM inset. Magnetization data were collected
while sweeping the magnetic field using sweep rates of about
300\,mT/min for both increasing and decreasing fields. Note that
due to hysteresis around the phase transitions observed at 1.8\,K,
the sweep rate was significantly varied in these field regions in
order to check for sweep-rate dependent effects. Using
quasi-static conditions, the observed small hysteresis in the
$M$($\mu_0H$) curves became negligible, as it is shown below.

For DC susceptibility and magnetization measurements down to
temperatures of 250\,mK an in-house-built cantilever magnetometer
was used, which works like a Faraday-force magnetometer. This
set-up was used to perform magnetization measurements in applied
magnetic fields up to 12\,T for $H\parallel b$ with a sweep rate
of 4\,mT/min.

\begin{table}[t!]
\caption{\label{tab:crystals} List of linarite crystals used for
the experiments presented in this work and former studies. This
study focuses on the following physical effects: Susceptibility
$\chi$, magnetization $M$, specific heat $C_p$, magnetocaloric
effect MCE, magnetostriction $\beta$, and thermal expansion
$\alpha$.}
\begin{ruledtabular}
\begin{tabular}{clll}
\# & origin & mass & methods\\
\hline
1 & Blue Bell Mine\footnotemark[1]& 26\,mg & NMR \cite{Wolter2012}, neutrons \cite{Willenberg2012}\\
2 & Blue Bell Mine& 6\,mg & $\chi$, $M$\\
3 & Blue Bell Mine& 205\,$\mu$g & $C_p$\\
4 & Blue Bell Mine& 0.98\,mg & $M$\footnotemark[3], MCE\\
5 & Blue Bell Mine& 11.62\,mg & $\alpha$, $\beta$\\
6 & Grand Reef Mine\footnotemark[2]& 6.22\,mg & $C_p$\footnotemark[4]\\
\end{tabular}
\end{ruledtabular}
\footnotetext[1]{Baker, San Bernadino, USA}
\footnotetext[2]{Graham County, USA} \footnotetext[3]{cantilever
mangetometer} \footnotetext[4]{high temperature data}
\end{table}

\begin{table*}[t]
\caption{\label{tab:structure} Structural parameters of linarite,
PbCuSO$_4$(OH)$_2$, at room temperature, as obtained from a
refinement of neutron scattering single-crystal data ($R_F = 100
\sum \left(\left|F_{\text{obs}}\right| - \sum
\left|F_{\text{calc}}\right|\right) / \sum
\left|F_{\text{obs}}\right|=6.7$, where $F$ represents the
structure factor). The thermal parameters $U_{ij}$ (given in
100\,\AA$^2$) are given in the form
$\exp\left[-2\pi^2\left(U_{11}h^{2}{a^*}^2+\ldots
2U_{13}hla^{*}c^{*}\right)\right]$. The thermal displacement of
sulfur was treated as isotropic since sulfur is a weak scatterer;
for details see text.}
\begin{ruledtabular}
\begin{tabular}{llllllllll}
 & $x/a$ & $y/b$ & $z/c$ & $U_{11}$ & $U_{22}$ & $U_{33}$ & $U_{12}$ & $U_{13}$ & $U_{23}$ \\ \hline
Pb & 0.3416(2)& 0.25& 0.3292(2)& 0.65(5) & 1.05(8) & 1.29(5) & 0 & -0.08(3) &0\\
Cu & 0& 0& 0 & 0.71(3) & 0.71 &0.71&0 &0 &0\\
S & 0.6692(4)& 0.25& 0.1159(6) & 0.47(6) & 0.47& 0.47&0 &0&0\\
O(1) & 0.5256(2)& 0.25& 0.9331(4)& 0.44(9)& 0.89(13)& 1.64(7)&0 &-0.014(51)&0 \\
O(2) & 0.6635(2)& 0.25& 0.4279(4) &1.93(10)&2.38(17)&0.77(6) &0 &0.58(6) &0\\
O(3) & 0.2535(1)& 0.5364(4)& 0.9420(3) &0.91(6) & 0.54(9)&2.12(5)& -0.26(7) & 0.29(3) &0.21(7)\\
O(4) & 0.9666(2)& 0.25& 0.7130(4) & 1.00(11) & 0.30(11) & 0.74(7) & 0 & 0.08(6) &0\\
O(5) & 0.0953(2)& 0.25& 0.2698(3)& 0.51(9) & 0.24(11) &0.90(7) & 0 & 0.01(6) & 0 \\
H(4) & 0.8667(4)& 0.25& 0.6166(8) & 1.48(19) & 1.84(25) &2.50(15) &0 & 0.11(12) &0 \\
H(5) & 0.0586(4)& 0.25& 0.4537(7) & 2.63(18) & 1.76(24) & 1.50(13) &0 & 0.52(11) & 0\\
\end{tabular}
\end{ruledtabular}
\end{table*}

\subsection{Specific heat and magnetocaloric effect}

Temperature-dependent specific-heat measurements at constant
magnetic fields along the $b$ direction have been performed using
a commercial cryostat system equipped with a 14\,T superconducting
magnet in combination with a homemade calorimeter providing a fast
relaxation measuring method \cite{Wang2001,Lortz2007}. The
heat-capacity platform is a modified $^3$He puck from the PPMS
setup (Quantum Design), the analyzing software is an in-house
development. The specific heat is continuously measured within one
large thermal relaxation step from $\Delta T + T_0$ to $T_0$, with
$\Delta T / T_0$ reaching up to 200\,\%. Here, $T_0$ is the bath
temperature and $\Delta T$ the temperature change during the
measurement. By using the temperature-dependent thermal
conductivity of our platform, we can calculate the specific heat
throughout this extended relaxation process, which takes about
60\,s. Compared to the conventional relaxation-time method this
technique allows for orders of magnitude faster data acquisition.
For the specific-heat measurements with magnetic fields applied
along $a_{\perp}$ and $c$ as well as for the zero-field
measurement up to 250\,K a commercial PPMS with a standard
measurement technique was used.

The magnetocaloric effect was measured for applied magnetic fields
up to 10\,T along the $b$ axis down to 300\,mK using an
in-house-built calorimeter. The temperatures of both the bath and
the sample were measured while sweeping the applied magnetic field
with a sweep rate of 75\,mT/min. The evolution of the temperature
difference arises from heating or cooling of the sample due to the
magnetocaloric effect.

\subsection{Magnetostriction and thermal expansion}

We have performed magnetostriction and thermal-expansion studies
using a capacitive dilatometer with a tilted-plate construction,
which is suitable for measurements parallel and perpendicular to
the magnetic field. The sample was placed in a cylindrical hole
between two round capacitance plates. In our case, we aligned the
$b$ axis parallel to the field and measured the length changes
along the $c$ axis. To determine absolute length changes, we have
calculated the corresponding capacitance changes by using a
capacitance bridge, Andeen-Hagerling AH2500A, with an effective
resolution of $10^{-5}$\,pF, which in our experiments corresponds
to minimal length changes of 1\,\AA. After subtracting the known
length change of the platform at a certain temperature and given
magnetic field it is thus possible to calculate  the absolute
length change of the sample as function of field or temperature.
The experiments have been carried out at temperatures ranging from
2 to 300\,K in fields up to 16\,T. The magnetostriction data  were
collected after stabilization of the temperature and using
quasi-static (sweep rate 0.3\,T/min) magnetic fields between 0 and
16\,T. The thermal expansion has been measured in constant
magnetic field using a temperature sweep rate of 0.2\,K/min.

\section{Results}

\subsection{Samples and diffraction}

So far, two sets of atomic positions were published for linarite
\cite{Schofield2009,Effenberger1987}, however these studies showed
a disagreement in the atomic $z$ coordinates. To determine an
accurate set of atomic positional parameters we performed
neutron-diffraction measurements using the D10 4-circle
diffractometer at the Institute Laue-Langevin within a recent
experimental study \cite{Willenberg2012}. 786 inequivalent nuclear
Bragg peaks were measured at room temperature using a neutron
wavelength of 1.26\,\AA. The structural parameters as obtained
from our refinement are listed in Tab. \ref{tab:structure}. This
way, we confirm the accuracy of the atomic coordinates published
by Effenberger {\it et al.} \cite{Effenberger1987} and present the
corresponding hydrogen positions.

\subsection{Susceptibility and magnetization}

\begin{figure}[t!]
\begin{center}
\includegraphics[width=0.9\columnwidth]{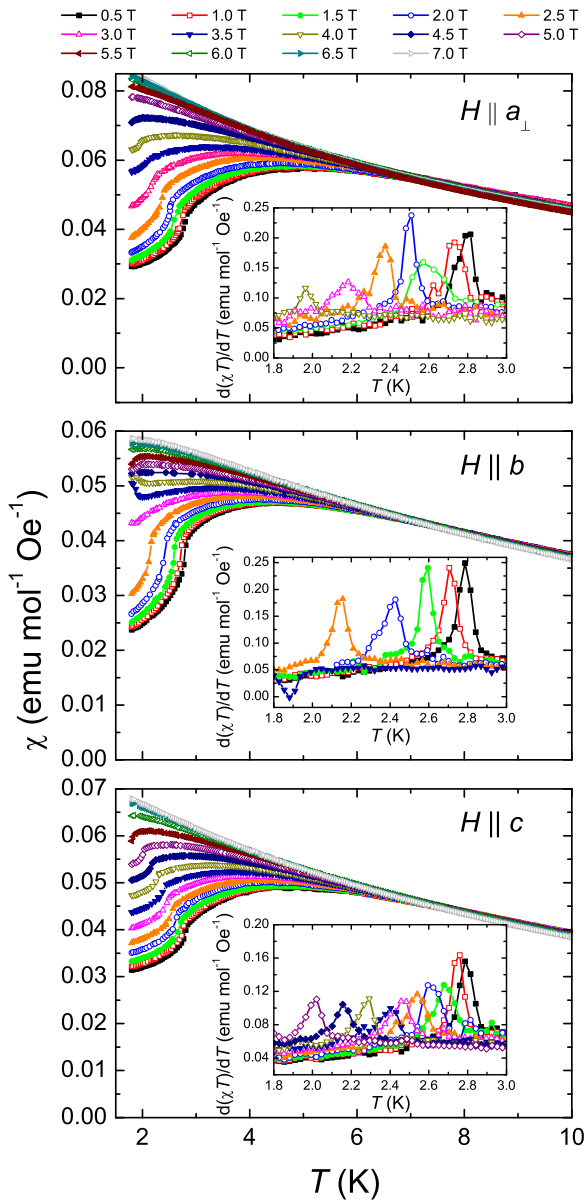}
\end{center}
\caption{(Color online) Susceptibility of PbCuSO$_4$(OH)$_2$ for
magnetic fields between 0.5 and 7\,T parallel to the
crystallographic $a_{\perp}$, $b$, and $c$ direction in the
temperature range from 1.8 and 10\,K. The insets depict the
temperature derivative of the product $\chi T$ for selected field
values used to determine the transition temperature $T_\text{N}$.}
\label{chi}
\end{figure}

In Fig.~\ref{chi}, we present the temperature dependence of the
macroscopic susceptibility of linarite for several magnetic fields
$H\parallel a_{\perp}$, $b$, and $c$, respectively. Here, the
susceptibility was measured in the temperature range from 1.8\,K
up to 10 K, while the magnetic field was varied from 0.5 to
7.0\,T. For small magnetic fields, the susceptibility has two
characteristic features: a broad maximum at around 5\,K and a
pronounced kink around 2.8\,K \cite{Wolter2012}. The maximum is
common to low-dimensional spin systems and is associated to
magnetic correlations within the Cu chains. Further, the kink
denotes a transition into a long-range magnetically ordered state
at the critical temperature $T_\text{N}$.

To determine the transition temperature as a function of the
magnetic field, the derivative d($\chi$$T$)/d$T$ has been
calculated for each field (see insets in Fig.~\ref{chi}). First,
we focus on the direction $H\parallel b$ because for this
direction the most remarkable physical properties, with a
multitude of field-induced phases, appear. To analyze the data it
is helpful to divide the measurements into three regimes: a
low-field region from 0--3.0\,T, an intermediate region from
$\sim$3.0--4.5\,T, and a high-field region from $\sim$4.5--7\,T.
The field dependence of the transition temperature $T_\text{N}$
differs from region to region. In the low-field region,
$T_\text{N}$ monotonously decreases with increasing field. In the
intermediate-field as well as in the high-field region the
transition temperature changes with different slopes. This
observation gives rise to the assumption that for this field
direction there appear to be three different distinct types of
magnetically ordered phases upon varying the magnetic field.

\begin{figure}[t]
\begin{center}
\includegraphics[width=0.9\columnwidth]{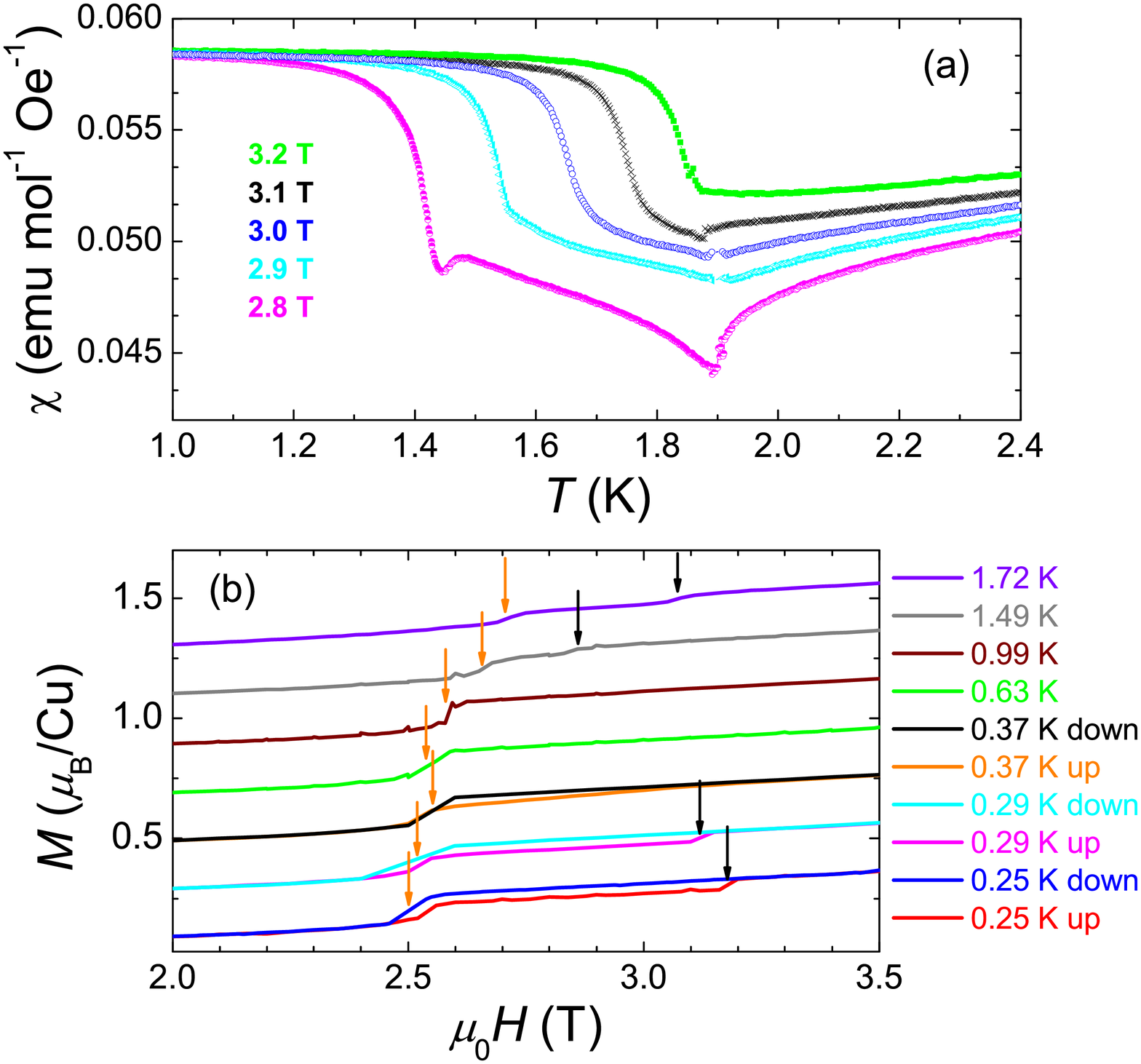}
\end{center}
\caption{(Color online) (a) Low-temperature susceptibility in
different fields for the intermediate-field range of
PbCuSO$_4$(OH)$_2$ for $H\parallel b$. (b) Field-dependent
magnetization of linarite for $H\parallel b$. The steps and
hystereses indicate field-induced transitions from the helical
ground state to another phase.  For clarity the curves are shifted
to each other.} \label{Mlow}
\end{figure}

In line with this argument, the susceptibility at low temperatures
in the low-field region shows an antiferromagnetic-like downturn,
while in the intermediate-field region an upturn, and in the
high-field region a downturn is observed. This suggests
qualitative changes regarding the types of magnetically ordered
phases present in linarite for magnetic fields directed along the
$b$ direction.

Furthermore, the susceptibility measured in the intermediate-field
region to lower temperatures and shown in Fig.~\ref{Mlow}(a)
displays two clear anomalies at 2.8\,T. With increasing magnetic
field the two transitions are pushed closer to each other, merging
at around 3.2\,T. Taken together, these observations clearly
justify the identification of three different magnetic phases.

\begin{figure}[t]
\begin{center}
\includegraphics[width=0.9\columnwidth]{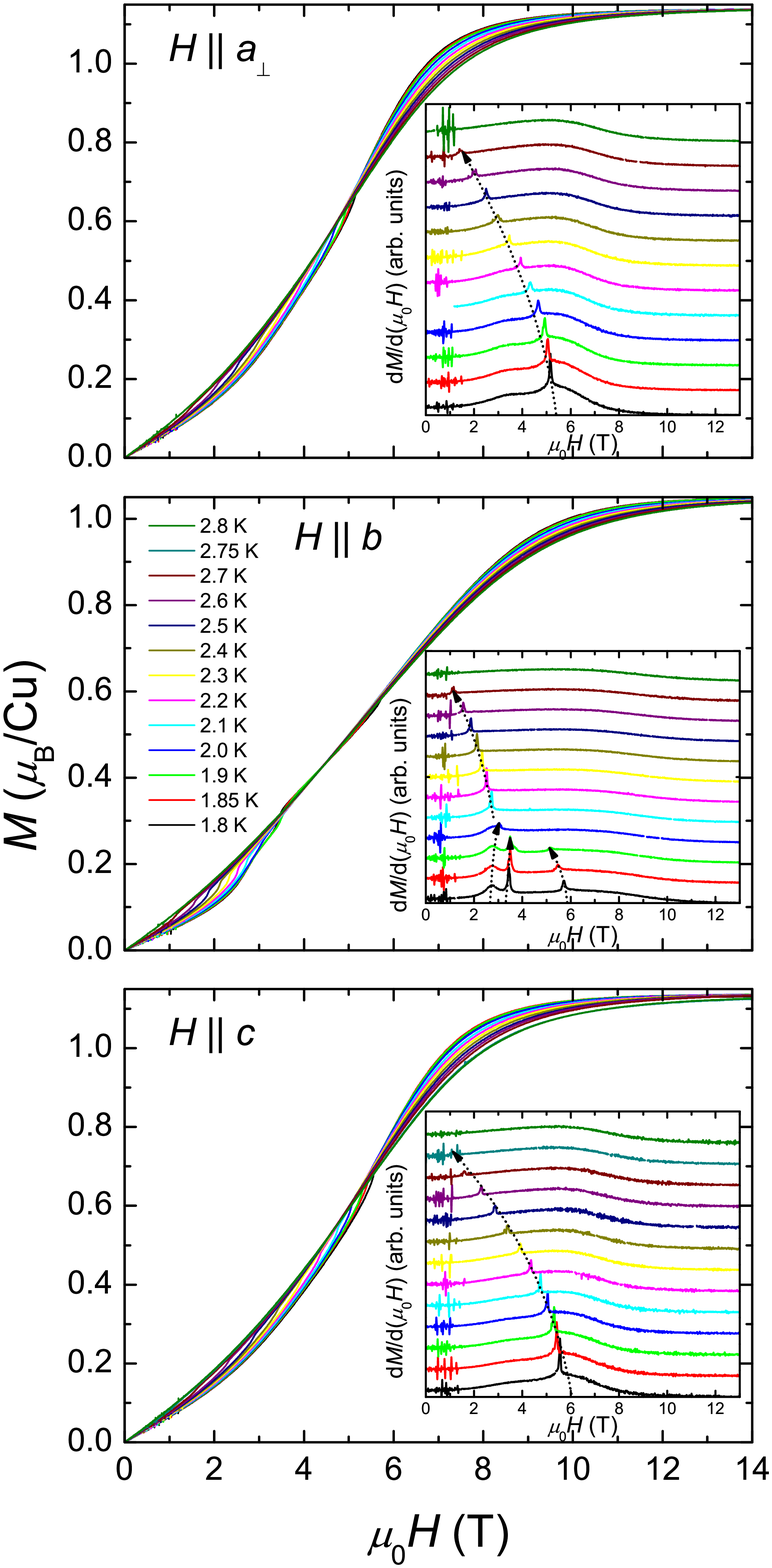}
\end{center}
\caption{(Color online) Magnetization data, $M(\mu_0H)$, and the
derivatives d$M$/d($\mu_0H)$ of PbCuSO$_4$(OH)$_2$ for all
crystallographic directions as a function of magnetic field in the
temperature range between 1.8 and 2.8\,K.} \label{M}
\end{figure}

In contrast, for magnetic fields aligned parallel to $a_{\perp}$
and $c$ the susceptibility behaves in a qualitatively similar
manner for all magnetic fields. For increasing magnetic fields,
the maximum in the susceptibility successively shifts to lower
temperatures, indicating a suppression of antiferromagnetic
fluctuations, which are gradually replaced by ferromagnetic
fluctuations. Moreover, only a monotonous decrease of the
magnetic-ordering temperature with increasing field is detected,
and the susceptibility always undergoes an antiferromagnetic-like
downturn at the transition. Consequently, for these field
directions the magnetically ordered phase basically corresponds to
the low-field phase for fields $H\parallel b$.

Next, in Fig.~\ref{M} we present the magnetization, $M(\mu_0H)$,
and the field derivatives d$M$/d($\mu_0H)$ of PbCuSO$_4$(OH)$_2$
as a function of field $H\parallel a_{\perp}$, $H\parallel b$, and
$H\parallel c$, respectively, for fixed temperatures between 1.8
and 2.8\,K. Measurements were carried out both for increasing and
decreasing field to check for hysteretic behavior. Altogether,
only a weak hysteresis was observed, depending on the field sweep
rate. For small sweep rates, viz., of the order of 0.1\,T/min, the
hysteresis is negligible. Therefore, here we only show the
up-sweep data using quasi-static measurement conditions at small
sweep rates.

As reported previously, a large anisotropic response is observed
in the saturation magnetization, $M_{\text{sat}}$, and in the
saturation field, $H_{\text{sat}}$ \cite{Wolter2012}. Here, we
focus on the anisotropy of the number of field-induced transitions
observed below 2.8\,K. Again, as for the susceptibility, the data
for $H\parallel a_{\perp}$ and $H\parallel c$ are similar and
differ from the data for $H\parallel b$. For $T < 2.0$\,K and
$H\parallel b$, there are three different peaks in the field
derivative d$M$/d($\mu_0H)$, i.e., at 1.8\,K at $\mu_0H_{c1}^{b}$
$\approx$ 2.7\,T,  $\mu_0H_{c2}^{b}$ $\approx$ 3.4\,T, and
$\mu_0H_{c3}^{b}$ $\approx$ 5.7\,T. With increasing temperature
the first transition shifts to higher fields and vanishes at
$\sim$2.1\,K. As well, the second transition shifts to higher
fields and vanishes at $\sim$2.0\,K, while the third transition
decreases in field and disappears at $\sim$2.0\,K. Next, a new
peak arises at 2.1\,K at about $\mu_0H_{c4}^{b} \approx 3.0$\,T,
which also decreases in field with increasing temperature and
fades out at $T_\text{N} \approx$ 2.8\,K.

In addition, from magnetization experiments for $H\parallel b$
down to 0.25\,K a two-step transition, i.e., two anomalies at
$H_{c1}^{b}$ and $H_{c2}^{b}$, has been studied. First, by
decreasing the temperature from 1.72\,K the double transition
associated to the intermediate-field phase transforms into a
single one at 0.99\,K [Fig.~\ref{Mlow}(b)]. Upon lowering the
temperature to less than 600\,mK, this intermediate-field regime
becomes hysteretic in the magnetization with respect to the
field-sweep direction. The transition/hysteretic region is defined
by steps in the magnetization indicated by the arrows in the
figure. The hysteretic region was also found by
magnetocaloric-effect measurements and will be discussed in more
detail in section \ref{ch:specific_heat}.

\begin{figure}
\begin{center}
\includegraphics[width=0.9\columnwidth]{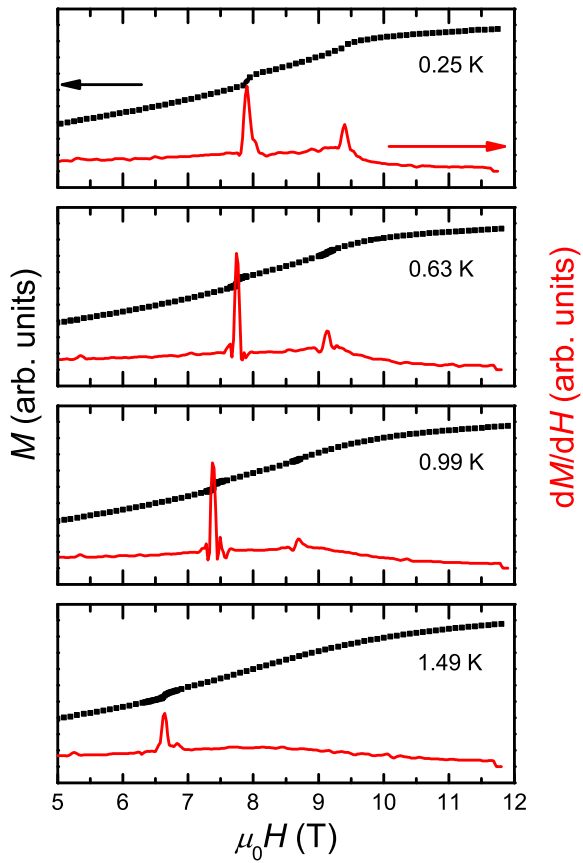}
\end{center}
\caption{(Color online) High-field magnetization and its field
derivative of PbCuSO$_4$(OH)$_2$ at low temperatures for
$H\parallel b$.} \label{Mhigh}
\end{figure}

The high-field/low-temperature magnetization data
(Fig.~\ref{Mhigh}) show that the shift of the third transition to
higher fields continues down to temperatures of 0.25\,K.
Furthermore, the data hint towards the existence of yet another
transition in fields of about 9\,T, as is indicated by a weak
feature in the field derivative of $M(\mu_0H)$ (Fig.~\ref{Mhigh}).

Altogether, the magnetization is in very good agreement with the
susceptibility, as again at least three different magnetic phases
are observed. In view of the recently discovered helical ground
state of linarite \cite{Willenberg2012}, the transitions at low
fields for $H\parallel b$, $H_{c1}^{b}$ and $H_{c2}^{b}$, could
possibly be associated to a spin-spiral reorientation process.
Moreover, the features in the magnetization might indicate
additional phase transitions or a first-order character of certain
transitions. Ultimately, neutron-scattering experiments in these
field-induced phases should shed light on these issues
\cite{Willenberg2014}.

For magnetic fields $H\parallel a_{\perp}$ and $H\parallel c$, the
derivative of the magnetization only shows one transition, which
decreases in field with increasing temperature and vanishes at
$T_\text{N}$. This magnetic phase corresponds to the ground state
phase for $H\parallel b$.

\subsection{Specific heat and magnetocaloric effect}
\label{ch:specific_heat}

\begin{figure}[t]
\begin{center}
\includegraphics[width=0.9\columnwidth]{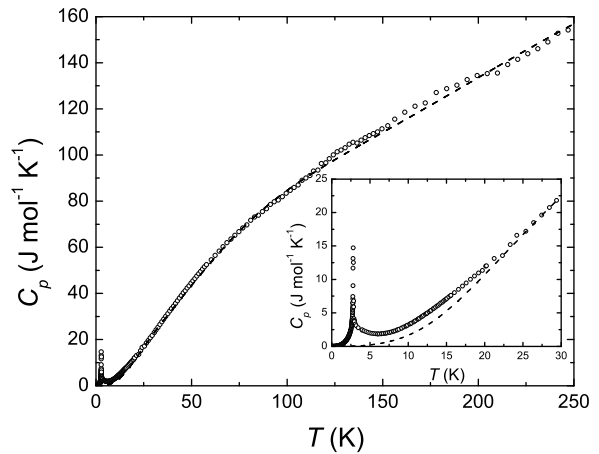}
\end{center}
\caption{Specific heat of linarite (sample \#6) in zero magnetic
field. The open circles represent the measured data, the dashed
line shows the modelled phononic contribution to the specific heat
(for details see text).} \label{cp0T}
\end{figure}

The specific heat, $C_p$, of PbCuSO$_4$(OH)$_2$ was measured in
magnetic fields up to 14\,T aligned along $a_{\perp}$, $b$, and
$c$ between 0.56 and 20\,K. Moreover, we also measured $C_p$ up to
250\,K in zero field (Fig.~\ref{cp0T}). The open circles represent
the measured specific heat, whereas the dotted line represents the
estimated phonon contribution, $C_{\text{ph}}$, to the specific
heat. The sharp peak in $C_p$ at 2.77\,K indicates the transition
into the long-range ordered magnetic state. Further, a fit
$C_{\text{ph}} \propto T^3$ does not produce the correct lattice
contribution above the transition temperature, since in the
temperature range up to $\sim$50\,K magnetic fluctuations are
present \cite{Wolter2012}. Therefore, as a first approximation a
simple harmonic model is developed to parameterize the phononic
specific heat using one Debye and two Einstein temperatures.

Linarite has 11 atoms per elemental formula unit, which implies
that 33 vibrational modes to the phononic specific heat exist.
Taking into account this constraint, we approximate the lattice
contribution to the specific heat by modelling it using one Debye
contribution together with two distinct Einstein terms. In
Fig.~\ref{cp0T}, we include the lattice contribution parameterized
by using 6 Debye modes with a Debye temperature of
$\mathnormal{\Theta_{\text{D}}} = 133$\,K, 9 Einstein modes with
an Einstein temperature $\mathnormal{\Theta_{\text{E,1}}} =
292$\,K and another 18 Einstein modes with
$\mathnormal{\Theta_{\text{E,2}}} = 1050$\,K.

\begin{figure}[b!]
\begin{center}
\includegraphics[width=0.9\columnwidth]{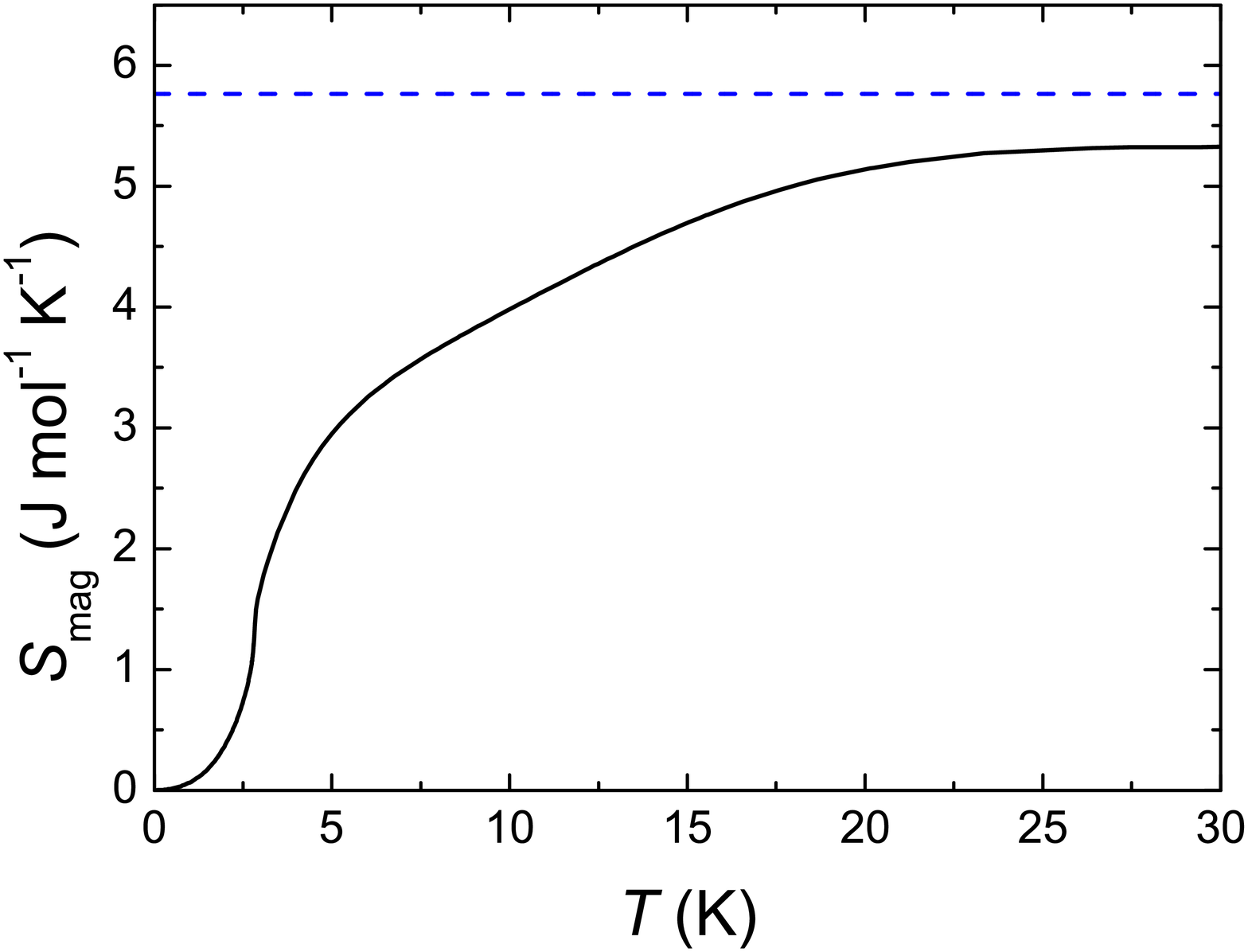}
\end{center}
\caption{Magnetic entropy of PbCuSO$_4$(OH)$_2$ in zero magnetic
field. The dashed line corresponds to the expected entropy for a
spin-$\tfrac{1}{2}$ system, $R\ln(2)$, while the solid line
indicates the entropy derived from the measured specific-heat
data.} \label{entropy}
\end{figure}

This parameterization of the lattice specific heat in principle
would need an experimental verification by means of for instance
inelastic neutron scattering. Most importantly, the obtained key
results are not influenced by subtleties in the choice of the
modelled lattice contribution, i.e., by the number of Debye and
Einstein contributions or by the used absolute values within
reasonable error bars. The used parameterization certainly will
oversimplify the phonon spectrum, a fact that needs to be taken
into account when comparing the experimental specific heat with
our theoretical modelling (see below). However, the values derived
for $\mathnormal{\Theta_{\text{D}}}$ and
$\mathnormal{\Theta_{\text{E}}}$ can be discussed on a qualitative
level. Especially, the Debye-like behavior of the lattice specific
heat with a rather low value $\mathnormal{\Theta_{\text{D}}} =
133$\,K is noteworthy in particular in the context of
multiferroicity, as it might possibly indicate a significant
magneto-elastic coupling in linarite.

Using the lattice contribution to the specific heat,
$C_\text{ph}$, derived this way, we proceed by determining the
magnetic part of the specific heat, $C_{\text{mag}} = C_p -
C_{\text{ph}}$. Next, we evaluate the entropy of
PbCuSO$_4$(OH)$_2$ associated with the magnetic contribution in
zero magnetic field by calculating the magnetic entropy
$S_{\text{mag}}$,
\begin{equation}\label{Smag}
S_{\text{mag}}(T) = \int_{0}^{T} \frac{C_{\text{mag}}}{T}
\mathrm{d}T,
\end{equation}
and which is depicted in Fig.~\ref{entropy}. Here, a total
magnetic entropy of $S_{\text{mag}} = R \ln(2J+1) = R \ln(2) =
5.76$\,J\,mol$^{-1}$\,K$^{-1}$ for Cu spin-$\tfrac{1}{2}$ spins is
expected. Experimentally, we obtain $S_{\text{mag}} =
5.32$\,J\,mol$^{-1}$\,K$^{-1}$ at $\sim$29.5\,K, which is in good
agreement with the expectation. This observation represents a
consistency check for our estimate of the phonon contribution.

\begin{figure}[t]
\begin{center}
\includegraphics[width=0.9\columnwidth]{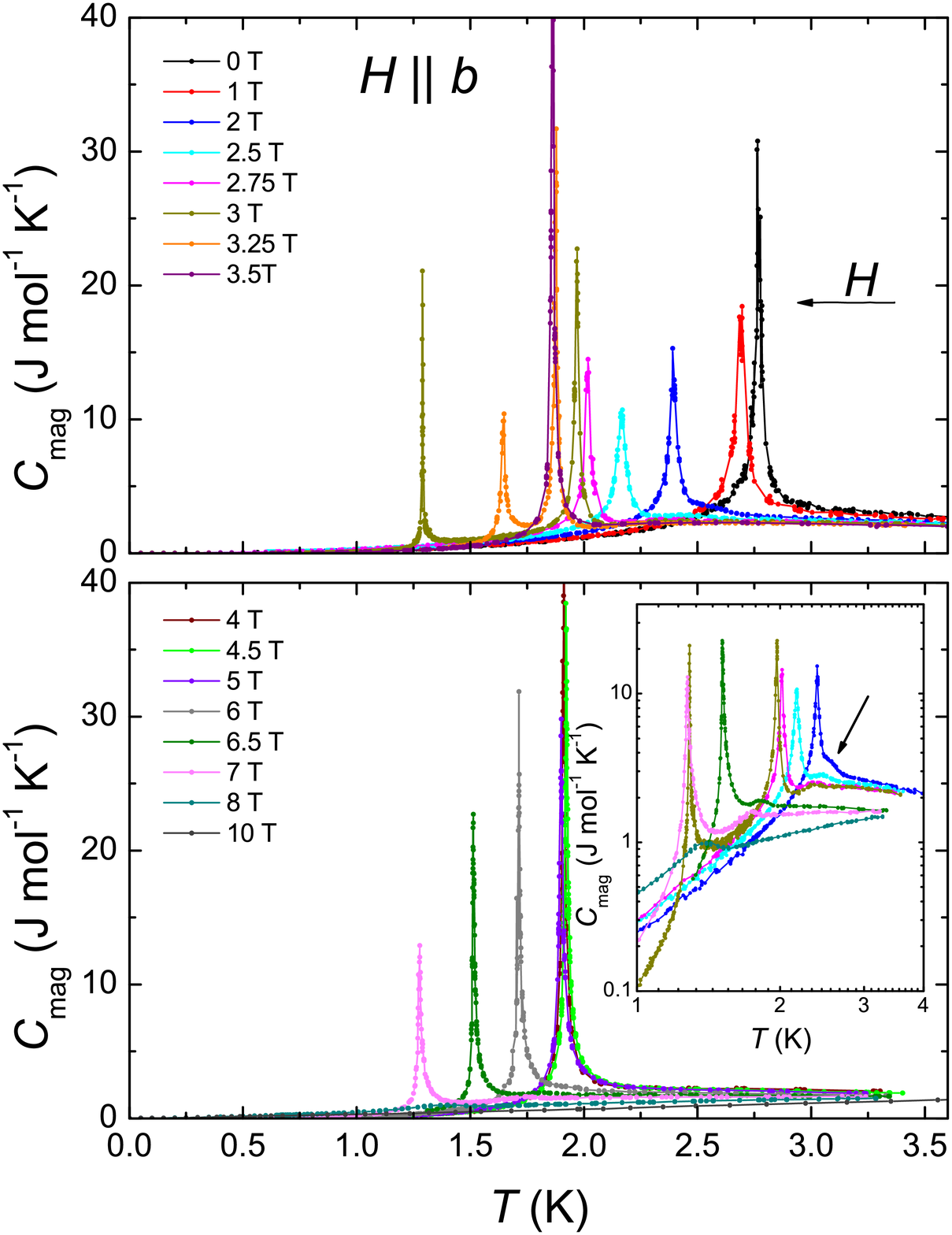}
\end{center}
\caption{(Color online) Magnetic specific heat of linarite (sample
\#3) as a function of the magnetic field aligned parallel to $b$.
The inset shows data at selected fields on a double-logarithmic
scale. The arrow indicates one of the many small anomalies that
hint towards another phase transition.} \label{heat}
\end{figure}

Moreover, from the temperature dependence of $S_{\text{mag}}$ we
find that down to $T_\text{N}$ there is a remarkable reduction of
the entropy. About 75\,\% of the total magnetic entropy are
associated to fluctuations above the magnetic 3D ordering. Such
behavior reflects the magnetic low-dimensional character of
linarite, with the remaining entropy associated to short-range
order and/or quantum fluctuations appearing in the temperature
range from above $T_\text{N}$ to about $\sim$50\,K
\cite{Wolter2012}.

Further, in Fig.~\ref{heat} we show the lattice-corrected specific
heat for $H\parallel b$ in fields up to 10\,T. Here, the upper
plot shows the data from 0 to 3.5\,T, the lower one the data from
4 to 10\,T. From zero field to 2.75\,T, the transition temperature
decreases with increasing field, while at 3 and 3.25\,T an
additional peak appears indicating an additional phase transition.
At 4 and 4.5\,T the transition temperature starts to increase
again with field, while it decreases for even higher fields.
Furthermore, a hump-like anomaly just prior to this transition
into the long-range ordered state is clearly discernible in the
field range 2--3\,T and 6.5--8\,T (see inset of Fig.~\ref{heat},
showing a log-log plot of the data at selected magnetic fields
with an arrow to exemplify one transition point). This anomaly
appears also to be connected to magnetic correlations which we
will discuss below.

\begin{figure}[b]
\begin{center}
\includegraphics[width=0.9\columnwidth]{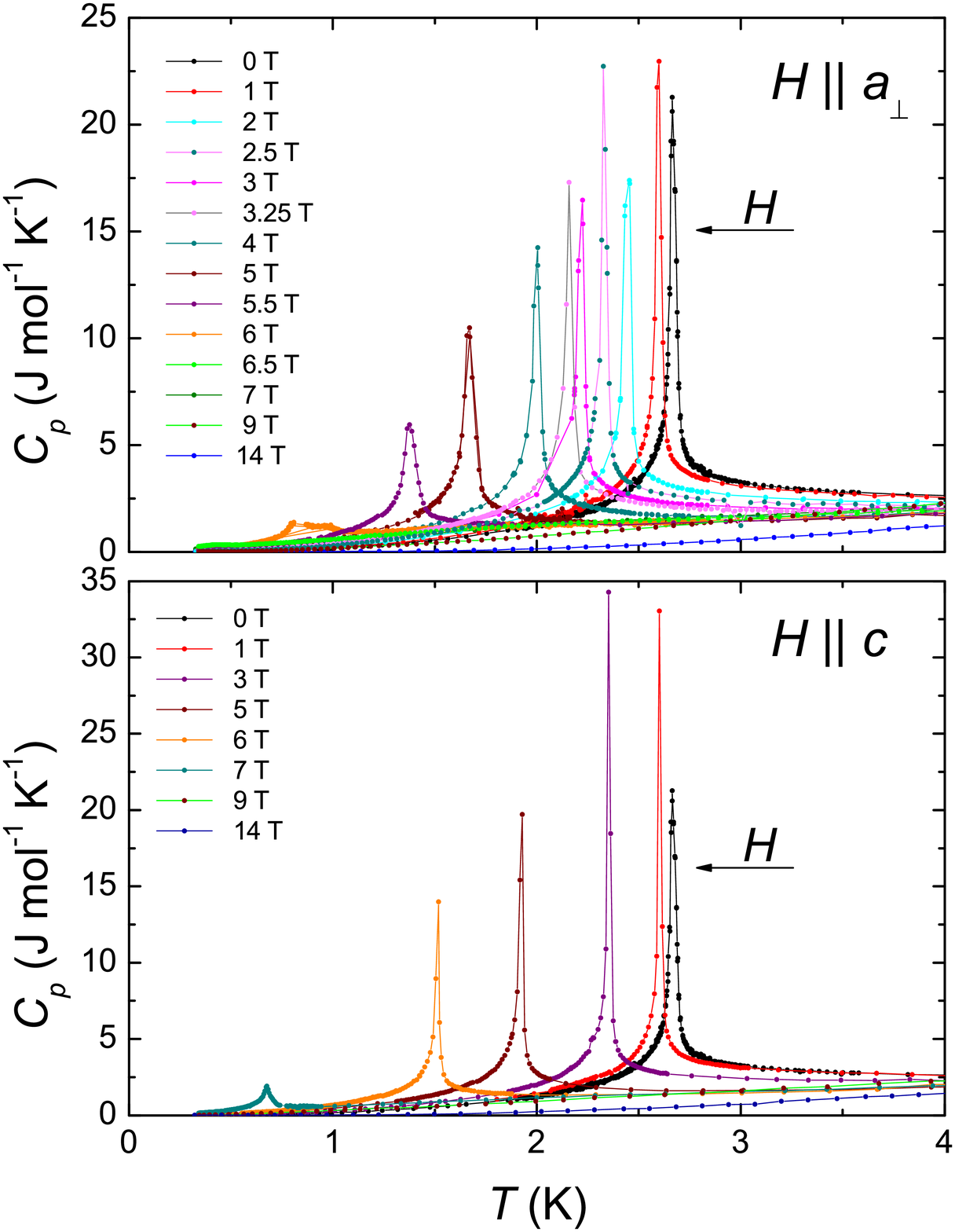}
\end{center}
\caption{(Color online) Specific heat of linarite (sample \#3) as
a function of magnetic fields aligned along $a_{\perp}$ and $c$.}
\label{all_Cp}
\end{figure}

For magnetic fields $H\parallel a_{\perp}$ and $H\parallel c$
(Fig.~\ref{all_Cp}), the specific heat shows only one sharp
anomaly, which is monotonously shifting to lower temperatures with
increasing magnetic field. This anomaly can be attributed to the
phase transition into the helical ground state.

\begin{figure}[t]
\begin{center}
\includegraphics[width=0.9\columnwidth]{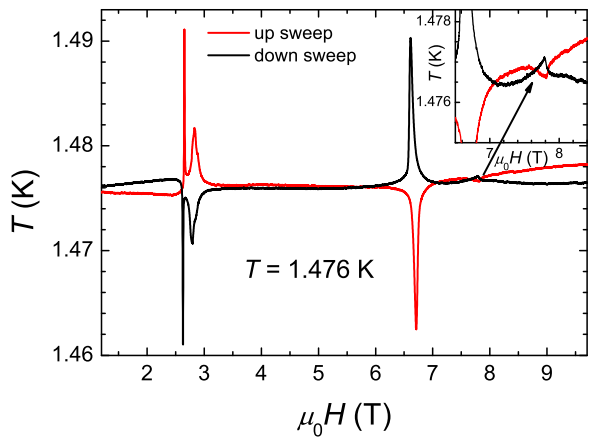}
\end{center}
\caption{(Color online) Field scan for the determination of the
magnetocaloric effect of linarite for $H\parallel b$ at a starting
temperature of 1.476\,K. The inset enlarges the feature seen in
the magnetocaloric effect in high magnetic fields.} \label{magcal}
\end{figure}

Next, in Fig.~\ref{magcal} we present a typical result of a field
scan in a magnetocaloric-effect measurement, here for a starting
temperature of 1.476\,K. In close resemblance to field scans for
the magnetization (Fig.~\ref{Mlow} and \ref{Mhigh}), various
transitions are visible at 2.65, 2.8, and 6.65\,T. The increase in
temperature of the up-sweeps and the decrease in temperature at
the down-sweeps at the first two transitions indicate that the
entropy is reduced above these transitions. In contrast, at the
third transition the entropy is increasing. Corresponding
experiments have been performed at various temperatures down to
0.3\,K (data not shown), allowing the determination of transition
fields analogous to those seen in the magnetization study.
Moreover, the inset of Fig.~\ref{magcal} enlarges the data at the
high-field region. As for to the magnetization experiment at high
fields and low temperatures a small feature appears (at about
7.8\,T), which shifts to higher fields and becomes more pronounced
with lowering the temperature. The fact that we observe features
both in the magnetization and in the magnetocaloric effect
indicates the existence of another phase transition.

Finally, the hysteretic phase at temperatures below $\sim$0.6\,K
and fields between 2.5\,T and 3.2\,T observed in the magnetization
was also investigated by means of the magnetocaloric effect (not
shown). Similar to the magnetization, pronounced and hysteretic
features have been observed here which can be associated with a
field-induced first-order phase transition.

\subsection{Magnetostriction and thermal expansion}
\label{ch:magnetostriction}

\begin{figure}[b]
\begin{center}
\includegraphics[width=0.9\columnwidth]{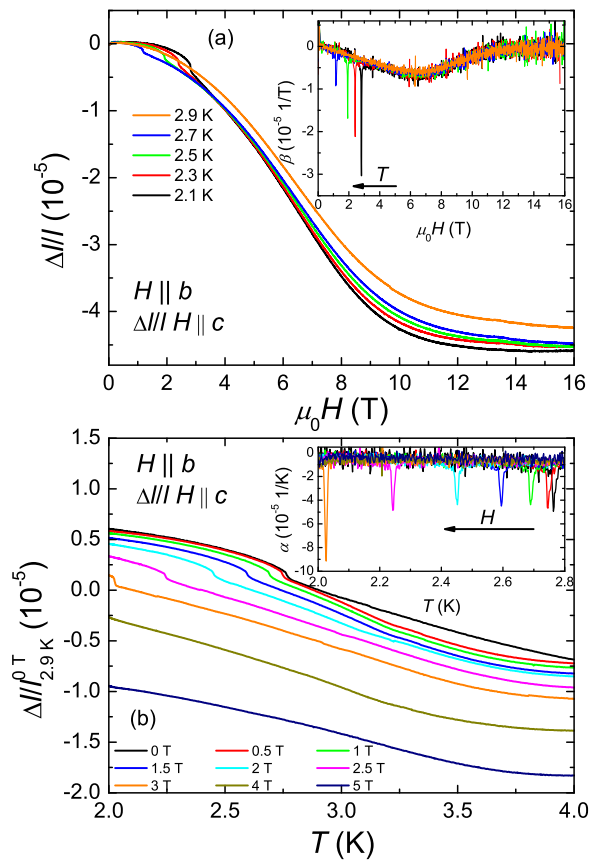}
\end{center}
\caption{(Color online) (a) Magnetostriction of linarite at
various temperatures as a function of magnetic field. (b) The
thermal expansion of linarite for various magnetic fields as a
function of temperature. The insets depict the field and
temperature derivatives $\beta$ and $\alpha$, respectively. Here,
the peaks indicate the transition into the long-range ordered
ground state.} \label{magstherm}
\end{figure}

In Fig.~\ref{magstherm}(a) and (b), we display the
magnetostriction and thermal-expansion data for magnetic fields
$H\parallel b$, respectively. For both experimental techniques the
length change of the sample was measured parallel to the $c$ axis
using sample \#5 in Tab. \ref{tab:crystals}, which has a length of
$\sim$0.95\,mm at room temperature. The magnetostriction was
measured at fixed temperatures between 2.9\,K and 2.1\,K while
varying the magnetic field from 0 up to 16\,T.
Fig.~\ref{magstherm}(a) shows the relative length change $\Delta
l/l$ as function of the magnetic field. Here, $l$ is the length of
the sample at room temperature and $\Delta l$ is the change of the
length due to the magnetostrictive effect.

For all measured temperatures the magnetostrictive effect is
negative with increasing magnetic field. Overall, after a strong
decrease of $\Delta l/l$ between 0 and 10\,T saturation sets in.
The transition into the long-range ordered state can be observed
as a downward step for temperatures up to 2.7\,K. The inset shows
the field derivative of the raw data,
\begin{equation}\label{beta}
\beta =\frac{\mathrm{d}}{\mathrm{d} (\mu_0 H)} \frac{\Delta l}{l},
\end{equation}
as a function of the magnetic field. The peaks in $\beta$
indicates the transition into the long-range ordered state,
shifting to lower magnetic fields upon increasing temperature. For
$T \geq 2.9$\,K no transition has been detected.

Next, in Fig.~\ref{magstherm}(b) the thermal-expansion data are
depicted. In this plot, the scale is defined by setting the length
change to zero at 2.9\,K and 0\,T, i.e., the scale is set by
$\Delta l/l_{2.9\,\text{K}}^{0\,\text{T}} = (l_{T}^{H} -
l_{2.9\,\text{K}}^{0\,\text{T}}) /
{l_{2.9\,\text{K}}^{0\,\text{T}}}$, in order to illustrate the
magnetostrictive effect. The data were obtained in the temperature
range from 2.0 to 4.0\,K in static magnetic fields up to 5\,T. For
all investigated magnetic fields, linarite shows a negative
thermal-expansion coefficient in the temperature range considered
here. The inset shows the derivative
\begin{equation}\label{alpha}
\alpha =\frac{\mathrm{d}}{\mathrm{d} T} \frac{\Delta
l}{l_{2.9\,\text{K}}^{0\,\text{T}}}
\end{equation}
as a function of temperature. Again, the transition temperature is
clearly seen as a sharp peak shifting to lower temperature upon
increasing magnetic field. For magnetic fields above 3.0\,T
magnetic long-range ordering occurs below 2.0\,K, which is below
the temperature range accessible with the present experimental
setup.

\section{Discussion}
\subsection{Magnetic phase diagram}
\begin{figure}
\begin{center}
\includegraphics[width=1\columnwidth]{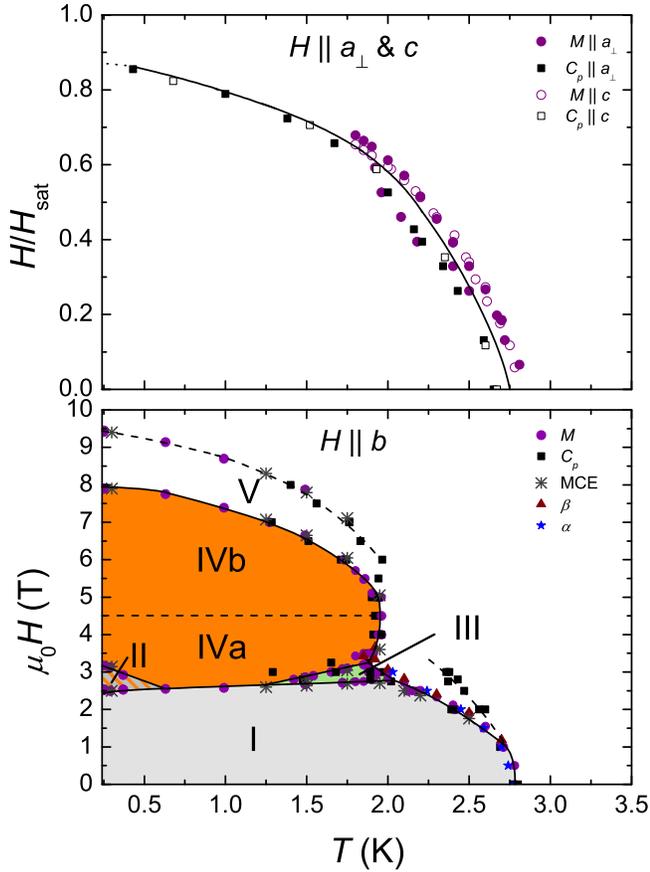}
\end{center}
\caption{(Color online) Magnetic phase diagram of
PbCuSO$_4$(OH)$_2$ for $H\parallel a_{\perp}$ and $c$ normalized
to $H_{\text{sat}}$ (upper panel) and for $H\parallel b$ (lower
panel) \cite{Willenberg2012}.} \label{phasediagram}
\end{figure}

From our experimental data, we derive the magnetic phase diagram
for linarite for fields $H\parallel a_{\perp}$, $b$, and $c$. The
lower part of Fig.~\ref{phasediagram} displays the phase diagram
for $H\parallel b$, which has already been presented in
Ref.~\cite{Willenberg2012}. Our experiments presented here give
evidence for five phases/regions in the phase diagram with
different physical properties:

\begin{description}
\item[Region I] \hfill \\
Region I represents the thermodynamic ground state of linarite,
with a helical magnetic order \cite{Willenberg2012} below 2.8\,K.
This phase is stable for fields up to about 2.7\,T at $T= 1.8$\,K
and about 3\,T at $T=2$\,K (see also the inset of Fig.~\ref{M}
(middle panel)). This phase boundary can be associated to a
spin-flop transition, generic for all CuO$_2$ chain compounds with
a rich phase diagram for the external field applied along the easy
axis.

The extrapolated spin-flop field $\mu_0H_{\text{SF}}(0)$ at $T=0$
according to the simplest possible phenomenological fit expression
\begin{equation}\label{eq:flop}
\mu_0 \left[ H_{\text{SF}}(T) - H_{\text{SF}}(0) \right] = A
T^\beta,
\end{equation}
yields $\mu_0H_{\text{SF}}(0) \approx 2.35(6)$\,T with $\beta
=0.61(15)$. This spin-flop field corresponds to a spin gap
$\Delta_\text{sg}=3.31$\,K or 0.289\,meV using $g_b=2.1$ derived
from our previous ESR data \cite {Wolter2012}. From
Eq.~\ref{eq:flop} we estimate 2.64\,T for $T=1.2$\,K. Also its
weak temperature dependence is rather remarkable: a sublinear
temperature dependence up to about 2.0\,K in our case as compared
to a subcubic dependence with $\beta =3.6$ in Li$_2$CuO$_2$ up to
5.5\,K \cite{Lorenz2011}. Noteworthy, both exponents differ from
the spin wave prediction $\propto T^{1.5}$ in leading order for a
classical unfrustrated  cubic antiferromagnet \cite{Feder1968}.

In the near future we plan a low-temperature ($< 1$\,K) ESR study
for linarite in order to check the value of the spin gap
$\Delta_{\rm sg}\approx 0.289$\,meV caused by the anisotropic
exchange estimated here from the spin-flop field and extrapolated
to $T=0$ (see Eq.~\ref{eq:flop}). We believe that the accurate
knowledge of $\Delta_\text{sg}$ provides a useful constraint for a
future refinement of the fundamental anisotropic interactions in
the very complex system under consideration as well as for a
phenomenological Landau-type free energy functional like in CuO
which is expected to be potentially useful for the description of
this and other monoclinic multiferroic systems
\cite{Villarreal2012,Quirion2013,Bogdanov2002} (for details see
next section).

\item[Region II] \hfill \\
Region II exists only at temperatures below $\sim$600\,mK, and is
defined by hysteresis effects in the magnetization and in the
magnetocaloric effect. It does possibly not represent a
thermodynamic phase, but a (possible first-order) crossover from
one phase to another.
\item[Region III] \hfill \\
The phase boundaries of phase III are possibly associated to
spin-spiral reorientation processes. Experimentally, we have
observed small discrepancies in the boundary positions from
measurements on samples from different origins. This indicates
that the sample quality/stoichiometry plays some role in this
phase. In turn, it reflects the frustrated nature of the magnetic
couplings in linarite, with the balance between different magnetic
phases being affected by variations of the local magnetic coupling
\cite{Willenberg2014}.
\item[Region IV] \hfill \\
Region IV can be divided into two regions, that is above and below
$\sim$4.5\,T, i.e., region IVa and IVb. While region IVa exhibits
a small additional ferromagnetic contribution in the temperature
dependence of the magnetic susceptibility at low temperatures,
region IVb instead shows an antiferromagnetic contribution. This
behavior, together with the pronounced anomalies in the specific
heat, suggests that in region IVa a long-range magnetically
ordered phase exists, where by canting of antiferromagnetically
aligned moments a small ferromagnetic signal is produced. Upon
increasing the field to above 4.5\,T this ferromagnetic signal is
saturated, resulting now in a predominantly antiferromagnetic
character of the susceptibility.
\item[Region V] \hfill \\
For region V, we find faint anomalies, i.e., a small hump-like
features in the specific heat, anomalies in the magnetocaloric
effect, and small jumps in the magnetization. The exact nature of
the magnetic ordering in region V, however, is unclear. Due to
those uncommon small features of the transition, we speculate that
short-range magnetic correlations play an important role in this
region.
\end{description}

Finally, the upper part of Fig.~\ref{phasediagram} depicts the
phase diagrams derived for fields aligned along $a_{\perp}$ and
$c$, respectively, plotted by normalizing the field to the
saturation field $H_{\text{sat}}$ for each direction, i.e.,
$H_{\text{sat}}^{a}=7.6$\,T and $H_{\text{sat}}^{c}=8.5$\,T
\cite{Wolter2012}. Here, for both directions only the helical
ground state phase of linarite is observed (region I for
$H\parallel b$). The scaling for both field directions attests the
close similarity of the phase diagrams for these geometries.

\subsection{Linarite in the context of frustrated chain
cuprates}

So far, about a dozen compounds have been assigned as quasi-1D $s
= \tfrac{1}{2}$ Heisenberg systems with competing ferromagnetic
nearest-neighbor and antiferromagnetic next-nearest-neighbor
intra-chain interactions. However, various fundamental issues such
as the existence of multipolar phases or the microscopic origin
for multiferroicity have not been comprehensively investigated up
to now. To set linarite into a proper context within this
challenging family of compounds, we will compare our observations
of its magnetic properties and the magnetic phases of linarite
with published reports for its magnetically analogous compounds.
As we will show, materials comparable to some extent to linarite
are LiCuVO$_4$ \cite{Enderle2005}, LiCuSbO$_4$ \cite{Dutton2012},
LiCu$_2$O$_2$ \cite{Masuda2004}, NaCu$_2$O$_2$ \cite{Capogna2005},
Li$_2$ZrCuO$_4$ \cite{Drechsler2007}, Li$_2$CuO$_2$
\cite{Mizuno1999}, CuO \cite{Quirion2013},
La$_6$Ca$_8$Cu$_{24}$O$_{41}$ \cite{Matsuda1996},
Ca$_2$Y$_2$Cu$_5$O$_{10}$ \cite{Matsuda1999}, CuGeO$_3$
\cite{Uchinokura2002}, Rb$_2$Cu$_2$Mo$_3$O$_{12}$ \cite{Hase2004},
Cu(ampy)Br$_2$ \cite{Kikuchi2000}, (N$_2$H$_5$)CuCl$_3$
\cite{Maeshima2003}, and Cu$_6$Ge$_6$O$_{18}$$\cdot x$H$_2$O
($x=0$ and 6) \cite{Hase2004a}.

In terms of the type of the magnetic ground state, LiCuVO$_4$,
LiCu$_2$O$_2$, NaCu$_2$O$_2$, Li$_2$ZrCuO$_4$, and CuO have the
most in common with linarite. They all exhibit a helically ordered
low-temperature phase, with LiCuVO$_4$
\cite{Gibson2004,Enderle2005,Schrettle2008,Svistov2011},
LiCu$_2$O$_2$
\cite{Masuda2004,Gippius2004,Gippius2006,Svistov2010,Bush2012},
and CuO \cite{Giovannetti2011,Villarreal2012,Quirion2013} showing
several field-induced phases. In Li$_2$ZrCuO$_4$, only a spin-flop
transition is observed \cite{Tarui2008}, while in NaCu$_2$O$_2$ no
significant changes of the magnetic properties in an external
magnetic field are registered
\cite{Capogna2005,Drechsler2006,Capogna2010,Leininger2010}.

Thus, the physical properties of LiCuVO$_4$, LiCu$_2$O$_2$, and
CuO are closest to those of linarite. In LiCuVO$_4$, the $\alpha$
value has been discussed controversially. LiCuVO$_4$ has been
described within a pure 1D model \cite{Enderle2010} using two
coupling constants, or alternatively by a 3D classical spin-wave
model \cite{Enderle2005} using 6 different $J$ values. However, if
one only compares the NN and NNN exchange, both models are in good
agreement with each other. Originally, Enderle {\it et al.}
\cite{Enderle2005,Enderle2010} proposed frustration ratios $5.5 >
\alpha > 1.42$, implying a concept of two weakly-coupled
antiferromagnetic chains. However, other authors arrived at
significantly different frustration ratios $\alpha \approx$
0.5--0.8
\cite{Sirker2010,Drechsler2011,Enderle2011,Nishimoto2012,Ren2012},
implying that a dominant ferromagnetic coupling prevails.
LiCuVO$_4$ undergoes long-range order below $T_\text{N} = 2.1$ K
into a spin-spiral ground state with a propagation vector ${\bf k}
= (0,\,0.532,\,0)$ and an isotropic ordered Cu$^{2+}$ moment of
0.31(1)$\mu_\text{B}$ \cite{Gibson2004}. The saturation field is
anisotropic and was determined as 52.1(3)\,T along the $b$ axis,
i.e., the chain direction, 52.4(2)\,T along $a$ and 44.4(3)\,T
along the $c$ axis \cite{Svistov2011}. LiCuVO$_4$ undergoes
transitions into different magnetic-field-induced phases for
fields aligned parallel to all crystallographic axes. It is argued
that at a critical field, $H_{c1}$, a spin-flop transition from
the spiral ground state occurs \cite{Buttgen2007,Nawa2013}. Based
on neutron diffraction \cite{Masuda2011} and NMR measurements
\cite{Buttgen2012,Nawa2013}, at a second critical field, $H_{c2}$,
a transition into a collinear spin-modulated structure is
proposed. However, this scenario is contested by recent neutron
scattering experiment, which is interpreted in terms of
quadrupolar correlations \cite{Mourigal2012}. Finally, at $H_{c3}$
a transition into a spin nematic phase has been proposed to occur
\cite{Zhitomirsky2010,Svistov2011,Mourigal2012}. For magnetic
fields along the $c$ axis, the phase boundary at $H_{c2}$ could
not be investigated so far, which is attributed to anisotropy
effects \cite{Banks2007,Buttgen2007,Schrettle2008,Svistov2011}.

In LiCu$_2$O$_2$, the magnetic exchange paths are still a matter
of debate. In Refs.~\cite{Masuda2004,Masuda2005}, a frustrated
double-chain system with large interchain interactions is favored
($\alpha = 0.54$). Conversely,
Refs.~\cite{Gippius2004,Maurice2012} support a scenario with
comparable values for the NN- and NNN-interactions ($\alpha
\approx 0.73$) and significantly smaller interchain interactions,
leading to a frustrated single-chain derived compound with
significant interchain coupling in the basal plane. LiCu$_2$O$_2$
undergoes a two-stage transition into a long-range ordered state
below $T_{c1} = 24.6$\,K and $T_{c2} = 23.2$\,K
\cite{Roessli2001,Zvyagin2002}. An incommensurate magnetic ground
state with a propagation vector ${\bf k} = (0.5,\,0.174,\,0)$ has
been established \cite{Masuda2004}, whereas the spin arrangement
could not be resolved so far. Masuda {\it et al.}
\cite{Masuda2004} favor a cycloidal spiral modulation along the
chain direction with spin spirals lying in the $ab$ plane. Park
{\it et al.} \cite{Park2007} suggest a spin spiral propagating in
the $bc$ plane. Finally, Kobayashi {\it et al.}
\cite{Yasui2009,Kobayashi2009} describe the ground state by
assuming an ellipsoidal spin helix in the $ab$ plane with a
helical axis tilted by $\sim$45$^{\circ}$ from the $a$ or $b$
axis, a view supported by Zhao {\it et al.} \cite{Zhao2012}. The
saturation field is estimated to be $\sim$110\,T \cite{Bush2012}.

LiCu$_2$O$_2$ has four highly anisotropic ordered phases. For
magnetic fields applied along the $b$ axis, i.e., the chain
direction, all four different phases appear: The helical ground
state below $T_{c2}$ and a field induced, hysteretic phase above
$H_{c1}$ which is interpreted as a spin-flop transition showing
pronounced sample dependencies \cite{Svistov2009,Bush2012}. On the
other hand, in Ref.~\cite{Sadykov2012} the absence of a sharp
reorientation transition was instead interpreted in terms of a
gradual rotation of the spinning plane of the spiral. The
intermediate phase between $T_{c1}$ and $T_{c2}$ is ascribed to a
collinear, sinusoidal structure with the spin direction along the
$c$ axis \cite{Yasui2009,Kobayashi2009}. Above $H_{c2}$ (which is
less anisotropic) another field-induced phase appears and is
discussed in the context of a collinear spin-modulated phase
similar to that in LiCuVO$_4$. For fields aligned along the $c$
axis, the spin spiral changes the direction of its spinning plane,
viz., does not undergo a spin-flop transition but enters directly,
into the supposed collinear spin-modulated phase
\cite{Sadykov2012}. Along the $a$ axis, the intermediate ordered
phase between $T_{c1}$ and $T_{c2}$ is absent but the sequence of
the field-induced phases is similar to that for $H\parallel b$
\cite{Bush2012}.

In comparison to these cases, in linarite ($\alpha = 0.36$) the
ordered moment in the helical phase below $T_\text{N} \approx
2.8$\,K varies from 0.638$\mu_B$ in the $ac$ plane to
0.833$\mu_\text{B}$ along the $b$ direction, according to the
propagation vector ${\bf k} = (0,\,0.186,\,0.5)$ of the spiral
\cite{Willenberg2012}. The Hamiltonian used to model linarite so
far contains two $J$ values and yields better results if some
anisotropy is included \cite{Wolter2012}. The saturation field is
a factor of $\sim$5 (12) smaller than in LiCuVO$_4$
(LiCu$_2$O$_2$) and even more anisotropic. Linarite shows five
different magnetic field-induced regimes down to 250\,mK, but for
fields along the $b$ axis only. The advantage regarding linarite
as compared to LiCuVO$_4$ or LiCu$_2$O$_2$ is that all magnetic
phases can be accessed in field-dependent neutron-scattering
experiments, which allows a direct measurement of the nature of
the ordering in the high-field phases. In turn, linarite is an
ideal material for testing the scenarios also put forward to
describe the high-field phases in LiCuVO$_4$ and LiCu$_2$O$_2$ as
well as serves to refine the underlying commonly used isotropic
AFM Heisenberg Hamiltonian, e.g., by the inclusion of different
spin anisotropies.

On the other hand, from a theoretical point of view, the complex
magnetic phase diagram of CuO seems to be closely related to the
one of linarite. CuO contains a three-dimensional network of
alternately stacked edge-shared CuO$_2$ chains coupled directly by
their edges. As a result of that stacking, buckled corner-shared
CuO$_3$ chains with a large antiferrmoagnetic NN-exchange integral
are formed, too (see Fig.~1 in Ref.~\cite{Rocquefelte2012}).
Noteworthy, the behavior of CuO is somewhat similar to that of the
chains considered here for linarite, when the magnetic field is
applied along the easy axis (see Fig.~7 of
Ref.~\cite{Quirion2013}). CuO contains six phases among them two
spiral/chiral phases, denoted as AF2 and HF2 in
Ref.~\cite{Villarreal2012} with the spiral propagation along the
easy axis for the AF2 phase as in our case.

According to Ref.~\cite{Rocquefelte2012} the $J_1$ of CuO is
antiferromagnetic (at a relatively large Cu-O-Cu bond angle of
96$^\circ$) and the pitch of the spiral should be obtuse, i.e.,
$\pi/2 < \phi < \pi$ in contrast to the acute pitch of linarite.
In this case, no multimagnon bound states as low-lying excitations
are expected for CuO in sharp contrast to such a possibility left
still for linarite (see Sect. V). Also the large AFM interchain
coupling for the former would exclude multipolar phases even for
change of sign of the NN interaction to be predicted for high
pressure \cite{Rocquefelte2012}. However, the authors of
Ref.~\cite{Jin2012} stress the important role of the frustrating
NN and NNN intrachain couplings in the stabilization of the spiral
state. In general, the situation with respect to the assignment of
the numerous exchange couplings involved is still under debate
even in the isotropic approach
\cite{Filippetti2005,Jin2012,Giovannetti2011,Rocquefelte2011,Giovannetti2011a,Pradipto2012}.
With respect to the anisotropic exchange, to the best of our
knowledge first of all the importance of the antisymmetric
Dzyaloshinskii-Moriya coupling has been discussed
\cite{Toledano2011,Giovannetti2011,Pradipto2012} whereas the
symmetric anisotropic exchange has been supposed to be weaker
\cite{Toledano2011}. However, a dominant Dzyaloshinskii-Moriya
interaction would remove the observed spin gap (spin-flop)
\cite{Benyoussef1999} in contrast to the available experimental
data for CuO \cite{Villarreal2012,Quirion2013}. A more detailed
comparison of commonalities and differences of the two similar
magnetic phase diagrams of linarite and CuO is postponed to a
future publication.

\section{Theoretical Aspects}

In this section, we discuss some theoretical aspects of the
one-dimensional isotropic $J_1$-$J_2$ model and its
generalizations to include interchain coupling and exchange
anisotropy in the light of the parameter region suggested by the
experimental studies described above and in
Ref.~\cite{Wolter2012}. In particular, the effect of an external
magnetic field on the specific heat within 1D models will be
discussed. Thereby, the main aim is to understand to what extent
such simplified effective models are meaningful for the
interpretation of the experimental data reported here and to
provide an outlook for future generalizations, where it will be
necessary. Here we show also results without a direct one-to-one
correspondence to our experimental results. These theoretical data
are of interest for the community working in the field of
theoretical quantum magnetism. This concerns mainly the field
dependence of the magnetic specific heat of the isotropic 1D
$J_1$-$J_2$ model. To the best of our knowledge this problem has
not been studied systematically in the literature. In this
context, we admit that the present state of the art of theory for
a rigorous description at arbitrary external magnetic fields at
any finite temperature doesn't allow to answer the corresponding
question about the nature of the individual phases shown in the
phase diagram in Fig.~\ref{phasediagram}. At the moment for many
physical quantities reliable theoretical predictions can be done
for high magnetic fields which equal the saturation fields and at
$T=0$ or at very low temperature.

First, we consider the isotropic $J_1$-$J_2$ model. We apply two
techniques: (i) the exact diagonalization (CED) for relatively
large finite periodic rings with $N$ = 16, 18, 20, and 22 sites
formally valid for any temperature but still affected by
finite-size effects manifesting themselves for instance in
artificially small gaps leading to an incorrect description at
very low $T$ and (ii) the transfer matrix renormalization group
(TMRG) technique \cite{Wang1997,Shibata1997} which treats the
infinite-chain limit at not too low temperature. In the present
calculations this lower limit is given by $10^{-3} |J_{1}|$, i.e.,
of about 0.1\,K, still below the lowest available experimental
data at 0.25\,K and the theoretical results presented recently in
Ref.~\cite{Huang2012}.

\begin{figure}
\begin{center}
\includegraphics[width=0.95\columnwidth]{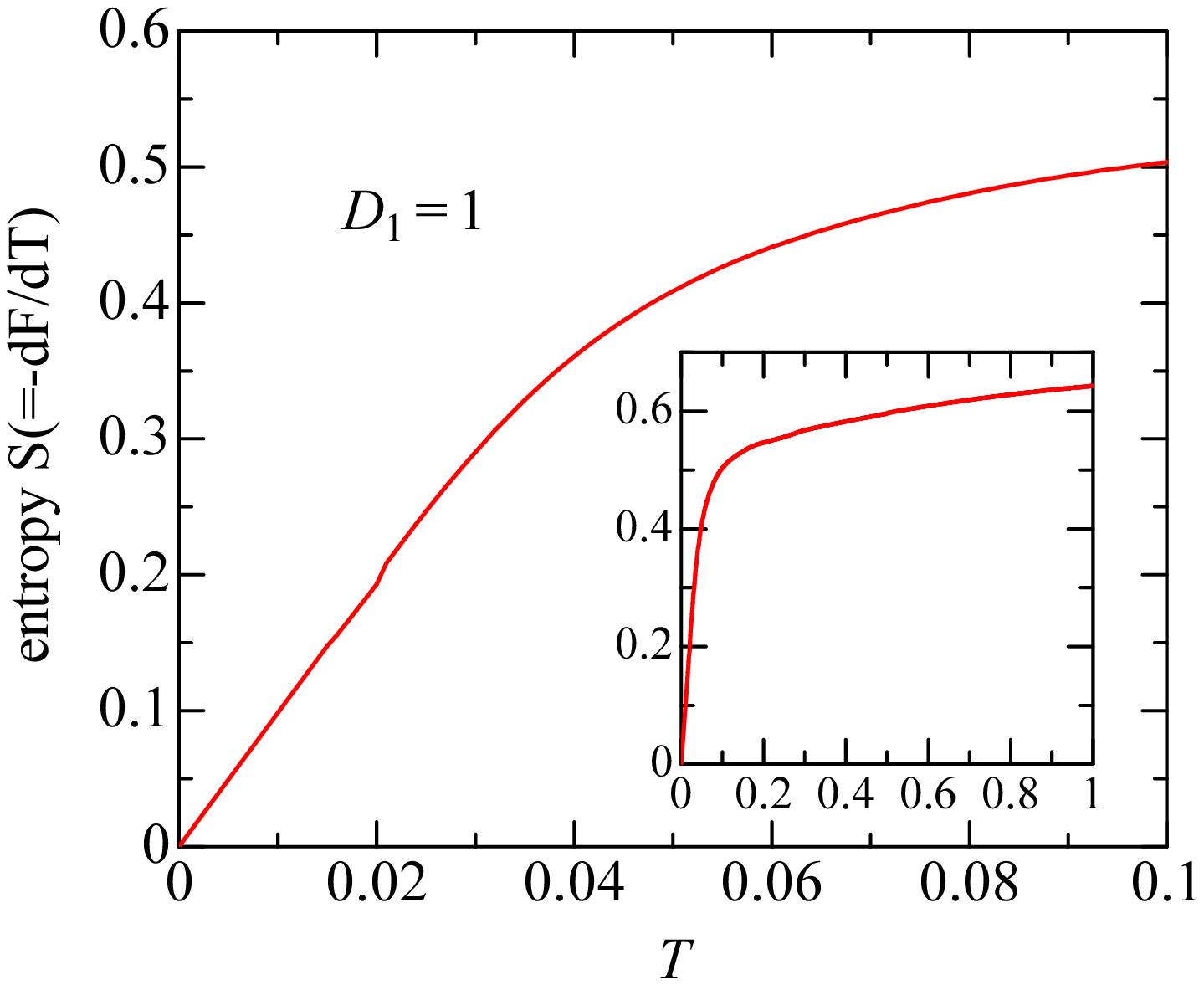}
\end{center}
\caption{(Color online) Low-temperature $T$-dependence of the
magnetic entropy for an 1D isotropic $J_1$-$J_2$ chain. The
temperature is measured in units of $|J_1|$. Inset: the same as in
the main figure on a larger temperature scale comparable with
$J_1$. The behavior for $T \rightarrow 0$ has been extrapolated
{\it linearly} to $T=0$ using the lowest available numerical TMRG
data (in between $T=0.006$ and 0.012) as suggested by the adopted
scenario of interacting spinons (see text).} \label{magentropy}
\end{figure}

\begin{figure}[b!]
\begin{center}
\includegraphics[width=0.95\columnwidth]{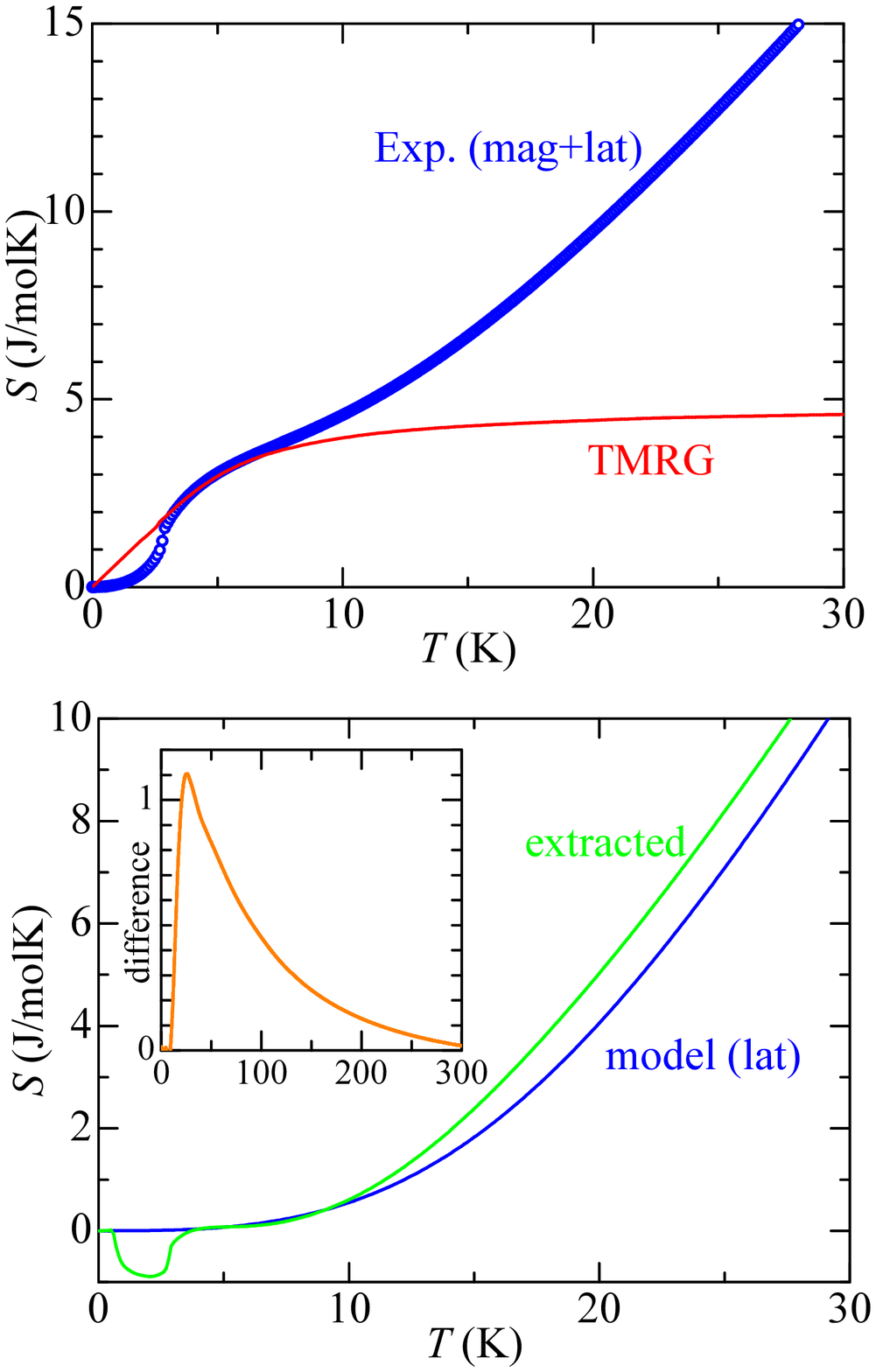}
\end{center}
\caption{(Color online) Upper panel: Temperature dependence of the
magnetic entropy for a 1D isotropic $J_1$-$J_2$ chain as compared
with the measured total entropy including the lattice
contribution. Lower panel: The phenomenological lattice
contribution resulting from a subtraction of the theoretical 1D
contribution shown in (a) from the measured total one as compared
with that from a harmonic-lattice model explained in the text.
Since the behavior of the theoretical curve for $T \rightarrow 0$
has been extrapolated linearly to $T=0$ (see also the note in the
caption of Fig.~\ref{magentropy}), the difference becomes
artificially negative in the region with magnetic ordering at
$T<T_\text{N}\approx 2.8$\,K where the 1D model naturally fails.
Inset: difference between the calculations and the above mentioned
harmonic model with one Debye spectrum and two Einstein modes.}
\label{comparisonentropy}
\end{figure}

In order to estimate the magnitude of the magnetic contribution to
the total specific heat and to evaluate the validity of the above
modelled harmonic lattice contribution, we start with the
calculated temperature dependence (in units of $|J_1|$) of the magnetic
entropy shown in Fig.~\ref{magentropy}. Adopting $J_1 = -94$\,K
and $\alpha = 0.36$ derived from our previous susceptibility fits
\cite{Wolter2012}, we arrive at $S \approx 0.55$ at $\sim$20\,K
which is still far from the high-$T$ saturation limit $\ln(2)
\approx 0.693$ (in units of $R$) or 5.76\,J\,mol$^{-1}$\,K$^{-1}$
in absolute units. Even at and slightly above 100\,K this value is
still by far not reached. Naturally, this behavior is more
pronounced for somewhat larger $|J_1|$ values which provide a
reasonable description of the saturation field: $|J_1|\approx
118.5(65)$\,K, which is shown in the upper panel of
Fig.~\ref{comparisonentropy}. The exact value of $J_1$ plays no
essential role in these considerations. For a refined estimate the
reader is referred to the discussion of the saturation field given
below. Returning to the lower panel of
Fig.~\ref{comparisonentropy} we show our extracted empirical
lattice part, too. One realizes a good description above about
3\,K, i.e., slightly above the magnetic ordering temperature of
2.8\,K, and below about 10\,K.

The overestimation of the experimental entropy by the theoretical
curve (based on a single-chain approach) at low temperatures shown
at low temperatures in Fig.~\ref{comparisonentropy} is rather
natural, because a pure 1D system on the spiral side exhibits no
magnetic ordering and hence its entropy must exceed that of the
magnetically ordered system at $T \rightarrow 0$. If the picture
of interacting spinons (living on the legs of the equivalent
zigzag-ladder and interacting via $J_1$) might be applied for that
case, a linear specific heat $C =\gamma T$ and correspondingly
also a linear entropy $S = \gamma T$ can be expected in that
limit, whereas in the ordered case dimensionality dependent higher
power-laws (quadratic and cubic in 2D and 3D cases, respectively)
are expected, which cause a faster decrease of the entropy at
low-temperature \cite{remark10}.

In fact, the experimentally observed $T^3$ dependence below
$T_\text{N}$ (not shown) further confirms the expected 3D ordering
already deduced from previous neutron-diffraction data
\cite{Willenberg2012}. Thereby, the total cubic term below
$T_\text{N}$ is found to exceeds very much the Debye contribution
to the harmonic lattice term obtained from the fit at
$T>T_\text{N}$. Approaching the critical point, the
low-temperature maximum of the specific heat and the inflection
point below it are down shifted to $T=0$, and $\gamma$
monotonously increases. In our case for $\alpha = 0.36$, i.e.,
well above the critical point at $\alpha_c = 0.25$, a remarkably
strong renormalization of the Sommerfeld coefficient already of
the order of 30 as compared to the case of non-interacting ``leg"
spinons for $J_1=0$ (i.e.\ $\alpha = \infty $) can be estimated. A
more quantitative analysis of the $J_1$ effect will be considered
elsewhere.

Above 10\,K  systematic deviations occur which point to a more
soft and/or anharmonic lattice model. In fact, the zig-zag
structure of hydrogen pairs along the chain might be interpreted
as an ``anti-ferroelectric" pseudo-spin ordering of hydrogen
positions described within interacting double-well potentials. The
observation that the intrachain exchange interactions are strongly
dependent on the actual hydrogen positions points to a strong spin
pseudo-spin interaction. This situation is reminiscent of the case
of Li$_2$ZrCuO$_4$ \cite{Moskvin2013,Vavilova2009} and of CuCl$_2$
\cite{Schmitt2009} where the pseudo-spin in the former case
results from the much heavier Li ions. In the present case of
light hydrogens even much stronger quantum effects might be
expected. In such a case the subdivision into a magnetic and a
lattice part might be difficult in general. However, below 10\,K
just where the low-$T$ maximum occurs at least a qualitatively
correct description might still be expected. The behavior below
3\,K seems to be dominated by the interchain coupling ignored in
this simple calculation.

\begin{figure}[b!]
\begin{center}
\includegraphics[width=0.8\columnwidth]{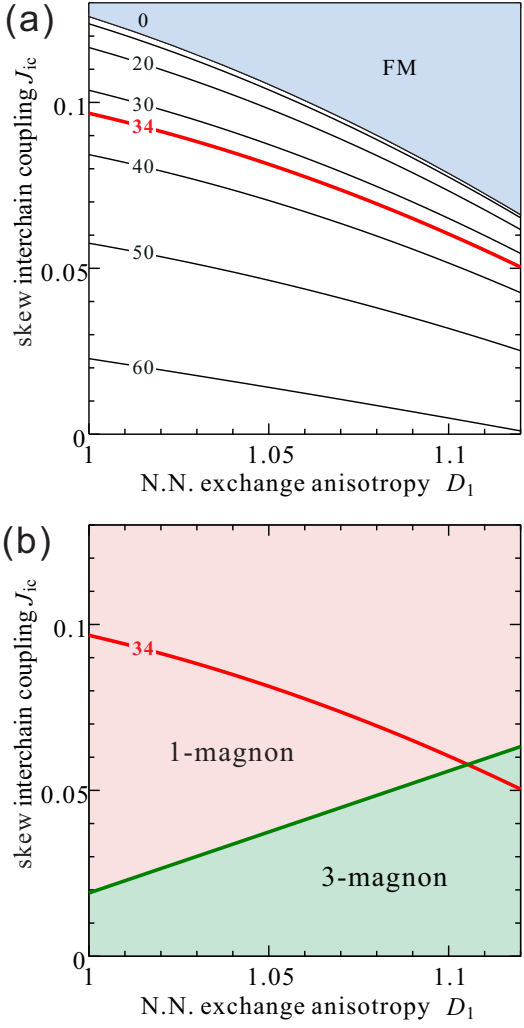}
\end{center}
\caption{(Color online) Influence of the interchain coupling
$J_\text{ic}$ and the easy-axis  exchange (spin) anisotropy $D_1$
of the ferromagnetic (FM) inchain NN-coupling $J_1$ on the ground
state of a system of coupled anisotropic $J_1$-$J_2$ spin chains,
cf. Eq.~\ref{Hamiltonian}, for an intrachain frustration rate
$\alpha=-J_2/J_1=0.36$. Upper panel (a): Zero-field plot of the
interchain coupling $J_\text{ic}$ vs. easy-axis  anisotropy $D_1$
for various  fixed pitch angles $\phi$ (given in degrees at the
left side of each curve). The FM ground-state phase  (i.e., $\phi
=0 $), present for large enough $J_\text{ic}$, is shown in the
light blue upper part of the figure. The NNN-coupling $J_2$ is
isotropic (i.e., $D_2=1$). Note that the red curve corresponds to
the observed pitch for linarite. Lower panel (b): Character of the
lowest excitations above the FM state for large  external field
above the saturation field applied in the easy-axis (b) direction.
These two figures, which have been slightly modified for clarity
here, are taken from S.~Nishimoto {\it et al.}
\cite{Nishimoto2013}.} \label{PD}
\end{figure}

\begin{figure}[b!]
\begin{center}
\includegraphics[width=0.95\columnwidth]{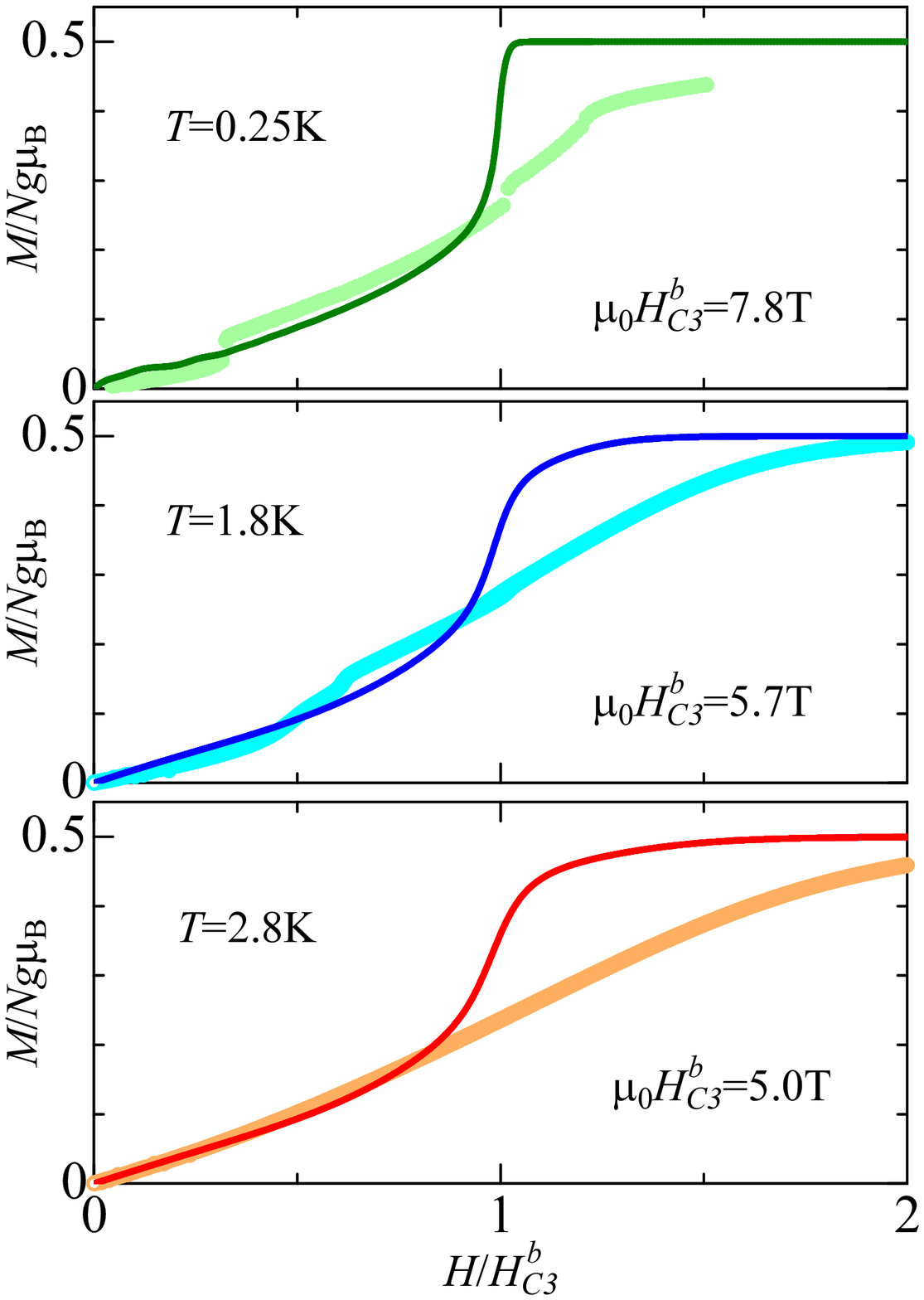}
\end{center}
\caption{(Color online) Magnetization at finite temperature for an
effective single chain  (1D) $J_1$-$J_2$ model for the frustration
ratio $\alpha=-J_2/J_1=0.365$ as compared with the experimental
data for $H\parallel b$. $H_{c3}$ is a fit parameter in order to
get a reasonable description at low fields. $H_{c3}$ corresponds
approximately to the inflection points of the experimental
magnetization curves shown in Figs.~\ref{M} and Fig.~\ref{Mhigh}.}
\label{magnetization}
\end{figure}

The experimentally obtained pitch angle $\phi$ for linarite of
about 33--34$^{\circ}$ is also strongly affected by the interchain
coupling and exchange anisotropy
\cite{Willenberg2012,Nishimoto2013}. From this, we estimate a 2D
saturation field of $9.5(6)$\,T at $T=0$, ignoring the very weak
interchain coupling in the third direction and taking $g=2.1$
derived from recent ESR data for the magnetic field parallel to
the $b$ axis. This number is in perfect agreement with the
experimental value of about 9.5\,T (see the upper panel of
Fig.~\ref{Mhigh}). Thereby, $J_1 = -112.6$\,K is the lower bound
for $J_1$ (taking into account the theoretical error bars $\pm
6.5$\,K from the $J_1$ estimate mentioned above). It has been
employed in order to minimize as much as possible the discrepancy
in the high-temperature entropy estimated from the applications of
the harmonic lattice model and of the theoretical approximation,
respectively. In the latter we used in addition to both 1D
couplings a skew (first diagonal) antiferromagnetic interchain
interaction of 5.6\,K and a 12\,\% easy-axis anisotropy for $J_1$
in order to have the correct pitch and a three-magnon phase for an
external magnetic field which equals the saturation field and
which is directed along the easy axis ($b$ axis) at $T=0$ (see
Fig. \ref{PD}). At such a field the system is fully
ferromagnetically polarized. Thereby, it is expected that a
similar diagram also holds for some slightly weaker fields and at
finite but low temperature. The latter $|J_1|$ values are slightly
smaller than the estimate given in our previous work for $|J_1|
\approx 138$\,K and a 10\,\% easy-axis anisotropy together with an
interchain coupling of 5.25\,K derived from the susceptibility
data (see Fig.~\ref{magnetization} in Ref.~\cite{Wolter2012}).
However, considering some uncertainty due to the final field value
in the susceptibility measurements and the approximate RPA
treatment of the interchain couplings in analyzing the 3D
$\chi(T)$ data, the direct estimate of $J_1$ from the measured
(extrapolated to $T=0$) saturation field is regarded to provide a
more accurate value.

The inspection of the interchain coupling vs.\ exchange anisotropy
``phase diagram" shown in Fig.~\ref{PD} clearly demonstrates that
a significant symmetric exchange anisotropy of at least of about
10\,\% is necessary to stabilize a multipolar-(octupolar) phase
(three-magnon bound phase). For details of the DMRG-based
calculations see Ref.~\cite{Nishimoto2010}. In this context it is
noteworthy that a significant exchange anisotropy suppresses
quantum fluctuations and this way contributes to the relatively
large magnetic moments observed in the spiral state (see
Ref.~\cite{Willenberg2012}) in spite of the pronounced quasi-1D
state with weak interchain coupling considered here. Thus, we may
conclude that a region near the top of the phase V or at very low
temperature in the experimental phase diagram shown in
Fig.~\ref{phasediagram} is in fact the place where one has still
some chance to detect such an exotic octupolar phase not yet
observed for any other real material to the best of our knowledge.
Further theoretical studies of even more complex spin chain models
and more detailed experimental studies are necessary to settle
this issue being of considerable theoretical interest.

Now, we reconsider the $T$ dependence of the magnetization for the
magnetic field $H\parallel b$ (see Fig.~\ref{magnetization}). The
inspection of Fig.~\ref{Mhigh} (upper panel) at low fields reveals
that the weak somewhat smeared kink in the experimental curve at
the lowest temperature of $T=0.25$\,K where data are available
corresponds approximately to the spin-flop field of about 2.46\,T
according to Eq.~\ref{eq:flop}. Let us now turn to an effective
isotropic 1D $J_1$-$J_2$ model. In general, a renormalization of
the effective $\alpha$ is expected due to the effects of the
interchain coupling and due to the easy-axis anisotropy present in
the material but ignored in our 1D model. Since the saturation
field is enhanced by the presence of antiferromagnetic interchain
interactions a smaller effective $\alpha$ than the more
``microscopic" one which enters a 2D or 3D model is expected in
order to compensate that enhancement. From the presence of the
easy-axis  anisotropy just the opposite is expected because it
lowers the saturation field, resulting in an overestimation of the
effective $\alpha$. Hence, the obtained effective
$\alpha_{\text{eff}} =0.365$ points to an approximate compensation
of both competing influences with a slightly larger effect from
the easy-axis anisotropy. The inspection of
Fig.~\ref{magnetization} demonstrates that only at high-fields
exceeding the saturation field a sizable $T$ dependence is
visible. The stronger deviations as compared to the hard-axis case
shown in our previous paper \cite{Wolter2012} points again to the
importance of anisotropy effects. In this context the presence of
antisymmetric contributions as given by the Dzyaloshinskii-Moriya
interaction (allowed by the low symmetry of the crystal structure
of linarite) may be assumed. However, the examination of such
interactions is beyond the scope of the present paper.

\begin{figure}
\begin{center}
\includegraphics[width=0.9\columnwidth]{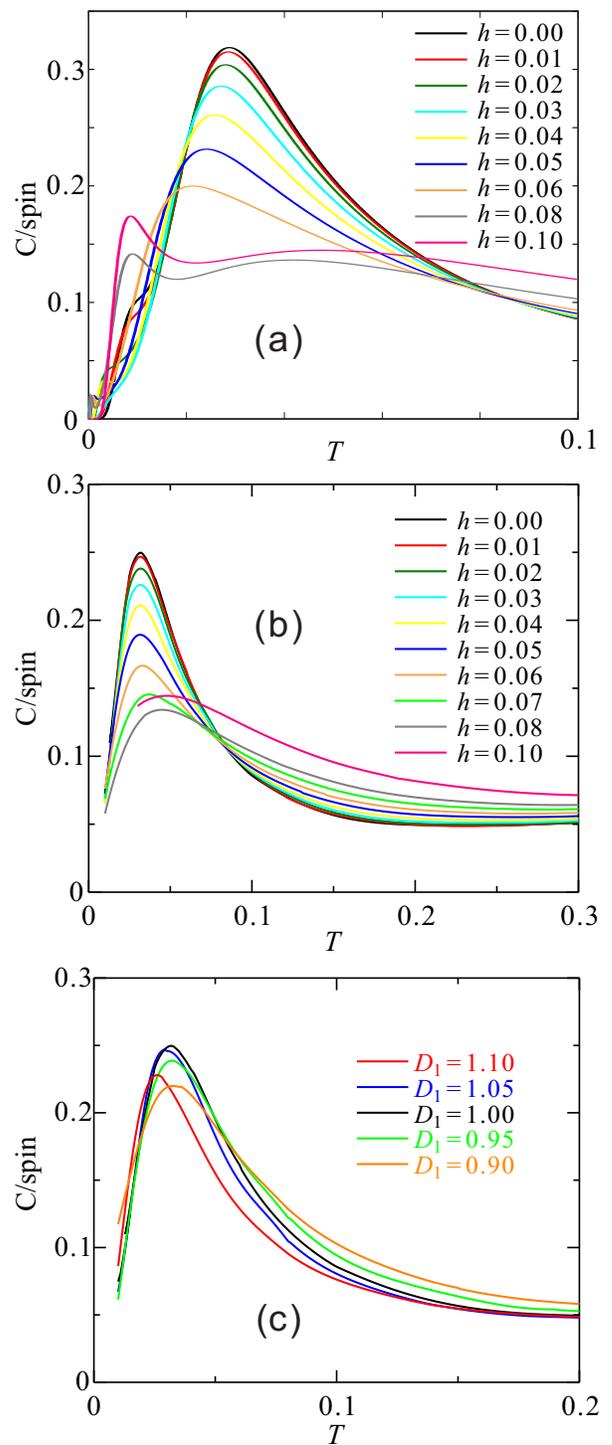}
\end{center}
\caption{(Color online) Temperature (in units of $|J_1|$)
dependence of the magnetic specific heat of a single chain within
the $J_1$-$J_2$ model. (a) Complete diagonalization-based
calculations for periodic rings with $N=22$ sites for different
dimensionless magnetic fields $h=gH/ |J_1|$. (b) The same as in
(a) for TMRG calculations. (c) $T$ dependence of the magnetic
specific heat at zero magnetic fields but with symmetric
anisotropic exchange included. $D_1 > 1$ means easy-axis
anisotropy for $J_1$, see Eq.~\ref{Hamiltonian}. $D_1 = 1$
corresponds to the isotropic limit of the $J_1$-$J_2$ model as
shown in (b), too}. \label{cvsh}
\end{figure}

Next, we consider the temperature dependence of the magnetic
specific heat at ambient  external magnetic fields. The results
are shown in Fig.~\ref{cvsh}. The zero-field magnetic specific
heat of the 1D $J_1$-$J_2$ model exhibits a well-known two-peak
structure (see e.g. Fig.~5 in Ref. \cite{Huang2012} for $\alpha =
0.4$) in a relatively broad region above \cite{Huang2012,Lu2006}
and below \cite{Haertel2008,Haertel2011} the critical point at
$\alpha= 0.25$. Thereby, for $\alpha > \alpha_{\text{c}}$ the peak
at low-temperature shifts towards $T=0$ approaching $\alpha_c $.
In the present case the high-temperature peak occurs near
0.66$|J_1|$ (not shown) whereas the low-temperature peak occurs
near 0.032$|J_1|$ within a pure 1D model, which corresponds to
about 3\,K for $|J_1| = 94$\,K mentioned above. Within the
anisotropic easy-axis model one finds a tiny down-shift up to
0.026$|J_1|$, i.e., to 2.9\,K assuming the larger $|J_1| \approx$
of 112.6\,K derived from the saturation fields discussed above
[see panel Fig.~\ref{cvsh}(c)].

The comparison of the behavior of the (3D solid) linarite with the
properties of the 1D models given above can be justified (at least
on a qualitative level) by a random phase approximation like
approach. Then such a correspondence is based on the knowledge
that a phase transition near the critical point due to finite
interchain coupling is triggered also by the sharp, well
pronounced low-temperature peak in the specific heat in the 1D
component. Thus, we have compared the somewhat broader peaks of
the 1D-models with the sharp peaks corresponding to the field
dependent phase transitions in the compound under considerations.
Experimentally, at ambient fields the magnetic phase transition
takes place at 2.8\,K. We ascribe that slightly smaller value as
compared to theoretical values of the peaks in the 1D models at
2.9~K and 3~K mentioned above to the effect of weak interchain
coupling ignored in both 1D approaches.

Finally, we summarize briefly the influence of the exchange
anisotropy on the magnetic specific heat [see Fig.~\ref{cvsh}(c)].
The account of a sizable easy-axis anisotropy for $J_1$ leads to a
down-shift of the low-temperature maximum and to a sharpening of
its peak. In the easy-plane case the opposite behavior is
observed. In both cases the discrepancy with the harmonic model is
not removed which suggests once again that the reason for the
discrepancy between an effective and the simple harmonic model is
not on the magnetic side but on the lattice model side.

We conclude this section with a critical comparison of both
theoretical methods we have employed to calculate the temperature
dependence of the magnetic specific heat. Considering the results
of our finite-cluster calculations using the spectrum obtained by
the CED depicted in Fig.~\ref{cvsh}(a), one realizes an observable
down-shift of the peak position down to 0.02 in fields from
ambient field to $h = 0.06$ (i.e. corresponding to about 5.6\,T
for $|J_1| \approx 119$\,K and $g = 2.1$, see the definition of
$h$ after Eq.~\ref{Hamiltonian} to be compared by a much smaller
shift obtained by the TMRG calculations). Anyhow, since
experimentally a much larger down-shift is observed for 7\,T,
only, we ascribe that difference to interchain coupling, too. The
study of that effect as well as the influence of various exchange
anisotropies is postponed to future studies in order to achieve a
better quantitative description of the experimental data. For
higher fields there is a clear upshift observed both in the CED
results for short rings and also within the TMRG (see panel (b)).
The appearance of further structures in the $C(T)$ curve
(including the second low-temperature peaks for the highest
fields, $ h = 0.08$ and $h = 0.1$) below $T \approx 0.02$ in the
CED data, see panel (a), is certainly  a finite-size artifact of
this approach.

\section{Summary}
In conclusion, we have determined the detailed magnetic phase
diagram of linarite by use of comprehensive thermodynamic
investigations. For magnetic fields aligned along the $b$
direction, linarite shows a rich variety of magnetic phases. This
phase diagram is even more complex than those of the related
frustrated spin-$\tfrac{1}{2}$ chain compounds LiCuVO$_4$ and
LiCu$_2$O$_2$. However, there are various similarities between the
different systems. We found remarkable similarities with the
magnetic phase diagram for the also monoclinic and multiferroic
CuO proposed very recently in the literature
\cite{Villarreal2012,Quirion2013} for the case of an external
magnetic field directed along the easy axis. A detailed and
comprehensive future comparison of both challenging systems is
expected to provide a deeper insight in the role of the frustrated
edge-shared CuO$_2$ chains in their crucial role for the rich
anisotropy effects observed here and there. In the case of
linarite, because of the relevant magnetic field scales,
neutron-scattering experiments will give a much deeper microscopic
insight into the magnetic phases and excitations of this material
as well as into this class of materials as a whole. Moreover,
based on our studies, linarite possibly is a candidate for showing
an octupolar (three-magnon) bound states hitherto experimentally
unknown. In addition, the expected highly anharmonic oscillatory
behavior of hydrogen points to the need for even more complex
models of strongly interacting spins and pseudo-spins as the
simplest model for the corresponding ferroelectric dipoles in the
extreme quantum limit (interacting two-level systems) for the
description of the quantum motion of hydrogen ions (protons) in
double- or multiple-well lattice potentials. To reach a deeper
understanding of these complex and challenging phases and
interactions further experimental and theoretical studies are
necessary.

\vspace{0.5cm}

\begin{acknowledgments}
We acknowledge fruitful discussions with N.~Shannon,
M.E.~Zhitomirsky, U.~R\"ossler, R.~Kuzian, J.~van den Brink, and
H.~Rosner as well as access to the experimental facilities of the
Laboratory for Magnetic Measurements (LaMMB) at HZB. We thank
G.~Heide and M.~G\"{a}belein from the Geoscientific Collection in
Freiberg for providing the linarite crystals \#1--5. S.-L.~D. and
J.~R. thank the Deutsche Forschungsgemeinschaft DFG for financial
support under the grants DR269/3-3 and RI615/16-3. This work has
partially been supported by the DFG under contracts WO1532/3-1 and
SU229/10-1.
\end{acknowledgments}


\bibliography{linarite_thermo}

\begin{thebibliography}{105}
\bibitem [{rem({\natexlab{b}})}]{remark10}%
  \BibitemOpen
  \href@noop {} {} ({\natexlab{b}}),\ \bibinfo {note} {{Adopting a specific
  heat $C_{\rm v,mag}=A_DT^D$ in the limit $T \rightarrow 0$, using
  Eq.~\ref{Smag} one has $S=(A_D/D)C_v$ for all dimensions $D=1,2,3$, i.e.,
  both quantities exhibit the same limiting dimensional dependent scaling
  behavior. These temperature dependencies of the magnetic specific
  $C_{v,\text{mag}}$ in that low temperature limit follow directly from the
  qualitatively well-known nature of the collective excitations and their
  power-law contribution to the magnetic specific heat. Indeed, in the ordered
  state in the 2D and 3D cases they are of bosonic character and can be well
  described within spin-wave theory by magnons with a linear dispersion law
  (near the $\Gamma$ point in the Brillouin zone) from spin-wave theory, just
  like acoustic phonons in the lattice part yielding a quadratic or cubic
  energy dependence of the bosonic density of states and a corresponding
  quadratic or cubic temperature of the free energy. Only in the disordered 1D
  case the excitations are given by spinons, i.e., fermions which results in a
  linear specific heat and entropy (in a formal sense magnons or confined
  spinons would also yield a linear law).}}\BibitemShut {Stop}%
\end{thebibliography}

\end{document}